\documentclass[final,1p,times]{elsarticle}

\usepackage{amssymb}

\usepackage[pagewise]{lineno}

\usepackage{amsmath}

\usepackage{epsfig}

\usepackage{algcompatible}
\usepackage{algorithm} 
\usepackage{algpseudocode}

\usepackage{graphicx}
\usepackage{epstopdf}

\usepackage[caption=false]{subfig}
\usepackage{balance}

\usepackage{array}

\usepackage{hyperref}

\journal{Parallel Computing}

\begin{document}

\begin{frontmatter}



\title{Speculative Segmented Sum for Sparse Matrix-Vector Multiplication on Heterogeneous Processors\tnoteref{labeltitlenote1}
\tnoteref{labeltitlenote2}
}


\tnotetext[labeltitlenote1]{The original title of this work presented at the 8th International Workshop on Parallel Matrix Algorithms and Applications (PMAA '14) was ``An Efficient and General Method for CSR-Format Based SpMV on GPUs''. For the present article, we improved our algorithm for emerging heterogeneous processors.}

\tnotetext[labeltitlenote2]{The source code of this work is downloadable at \url{https://github.com/bhSPARSE/Benchmark_SpMV_using_CSR}.}


\author[labeladd]{Weifeng Liu\corref{cor1}}
\cortext[cor1]{Corresponding author}
\ead{weifeng.liu@nbi.ku.dk}
\author[labeladd]{Brian Vinter}
\ead{vinter@nbi.ku.dk}
\address[labeladd]{Niels Bohr Institute, University of Copenhagen, Blegdamsvej 17, 2100 Copenhagen, Denmark}

\begin{abstract}

Sparse matrix-vector multiplication (SpMV) is a central building block for scientific software and graph applications. Recently, heterogeneous processors composed of different types of cores attracted much attention because of their flexible core configuration and high energy efficiency. In this paper, we propose a compressed sparse row (CSR) format based SpMV algorithm utilizing both types of cores in a CPU-GPU heterogeneous processor. We first speculatively execute segmented sum operations on the GPU part of a heterogeneous processor and generate a possibly incorrect results. Then the CPU part of the same chip is triggered to re-arrange the predicted partial sums for a correct resulting vector. On three heterogeneous processors from Intel, AMD and nVidia, using 20 sparse matrices as a benchmark suite, the experimental results show that our method obtains significant performance improvement over the best existing CSR-based SpMV algorithms.

\end{abstract}


\end{frontmatter}


\section{Introduction}
Sparse matrix-vector multiplication (SpMV) is perhaps the most widely-used non-trivial sparse basic linear algebra subprogram (BLAS) in computational science and modeling. The operation multiplies a sparse matrix $A$ of size $m\times n$ by a dense vector $x$ of size $n$ and gives a dense vector $y$ of size $m$. Despite its simplicity at the semantic level, an efficient SpMV implementation is generally hard, because $A$'s sparsity structure can be very irregular and unpredictable.

Compared to CPUs, co-processors (e.g., GPUs and Xeon Phi) promise much higher peak floating-point performance and memory bandwidth. Thus a lot of research has focused on accelerating SpMV on co-processors. One straightforward way on utilizing co-processors is to develop all-new sparse matrix formats (e.g., HYB~\cite{Bell:Implementing}, Cocktail~\cite{Su:clSpMV}, JAD~\cite{Li:GPU}, ESB~\cite{Liu:Efficient}, BCCOO~\cite{Yan:yaSpMV} and BRC~\cite{Ashari:An}) for specific hardware architectures. The experimental results showed that these formats can provide performance improvement for various SpMV benchmarks. 

However, the completely new formats bring several new problems. The first one is backward-compatibility. When the input data are stored in basic formats such as compressed sparse row (CSR), a format conversion is required for using the new format based SpMV. In practice, fusing a completely new format into well-established toolkits (e.g., PETSc~\cite{Balay:PETSc}) for scientific software is not a trivial task~\cite{Minden:Preliminary} because of the format conversion. Moreover, Kumbhar~\cite{Kumbhar:Performance} pointed out that once an application (in particular a non-linear solver) needs repeated format conversion after a fixed small number of iterations, the new formats may degrade overall performance. Furthermore, Langr and Tvrd\'{\i}k~\cite{Langr:Evaluation} demonstrated that isolated SpMV performance is insufficient to evaluate a new format. Thus more evaluation criteria, such as format conversion cost and memory footprint, must be taken into consideration. Secondly, when the SpMV operation is used with other sparse building blocks (e.g., sparse matrix-matrix multiplication~\cite{Liu:An}) that require basic storage formats, using the all-new formats is less feasible.

To leverage the SpMV performance and the practicality, Liu and Vinter proposed the CSR5 format~\cite{Liu:CSR5} to extend the basic CSR format. The experimental results showed that the CSR5 format delivers excellent SpMV performance, but merely needs very short format conversion time (a few SpMV operations) and very small extra memory footprint (around 2\% of the CSR data). Because the CSR5 format shares data with the CSR format, the CSR-based sparse BLAS routines can efficiently work with the CSR5 format. However, when a solver only needs a few iterations, the CSR5 may not deliver speedups, compared to using the basic CSR data.

Thereofore, improving performance of SpMV using the most widely supported CSR format has also gained plenty of attention~\cite{Bell:Implementing, Su:clSpMV, Williams:Optimization, Greathouse:Efficient, Ashari:Fast, Blelloch:Segmented, Harris:CUDPP, Baxter:Modern}. Most of the related work~\cite{Bell:Implementing, Su:clSpMV, Williams:Optimization, Greathouse:Efficient, Ashari:Fast, Liu:LightSpMV} has focused on improving row block method for the CSR-based SpMV. However, these newly proposed approaches are not highly efficient. The main reason is that co-processors are designed for load balanced high throughput computation, which is not naturally offered by the row pointer information of the CSR format. On the other hand, using segmented sum method for the CSR-based SpMV has been proposed by~\cite{Blelloch:Segmented} and been implemented in libraries cuDPP~\cite{Harris:CUDPP, Sengupta:Scan, Garland:Sparse} and Modern GPU~\cite{Baxter:Modern} for nVidia GPUs. Unlike the row block methods, the segmented sum algorithms evenly partition an input matrix $A$ for nearly perfect load balancing, and thus may be suitable for a co-processor implementation. But unfortunately, this method cannot recognize empty rows and requires more costly global operations. These extra overheads may offset performance gain of load balanced segmented sum and degrade overall SpMV efficiency.

Recently, heterogeneous processors (which are also known as heterogeneous chip multiprocessors) have been designed~\cite{Kumar:Heterogeneous, Keckler:GPUs} and implemented~\cite{Branover:Llano, AMD:Compute, Damaraju:22nm, nVidia:Tegra, Qualcomm:Snapdragon}. Compared to homogeneous processors such as CPUs or GPUs, heterogeneous processors can deliver improved overall performance and power efficiency~\cite{Chung:Single}, while sufficient heterogeneous parallelisms exist. The main characteristics of heterogeneous processors include unified shared memory and fast communication among different types of cores (e.g., CPU cores and GPU cores). Practically, heterogeneous system architecture (HSA)~\cite{HSA:Manual}, OpenCL~\cite{Munshi:The} and CUDA~\cite{Negrut:Unified} have supplied toolkits for programming heterogeneous processors.

Our work described in this paper particularly focuses on accelerating CSR-based SpMV on CPU-GPU heterogeneous processors. The main idea of our SpMV algorithm is first speculatively executing SpMV on a heterogeneous processor's GPU cores targeting high throughput computation, and then locally re-arranging resulting vectors by the CPU cores of the same chip for low-latency memory access. To achieve load balanced first step computation and to utilize both CPU and GPU cores, we improved the conventional segmented sum method by generating auxiliary information (e.g., segment descriptor) at runtime and recognizing empty rows on-the-fly. Compared to the row block methods for the CSR-based SpMV, our method delivers load balanced computation to achieve higher throughput. Compared with the classic segmented sum method for the CSR-based SpMV, our approach decreases the overhead of global synchronization and removes pre- and post-processing regarding empty rows.  

This paper makes the following contributions:

\begin{itemize}

  \item  We propose a fast CSR-based SpMV algorithm that efficiently utilizes different types of cores in emerging CPU-GPU heterogeneous processors. 

  \item  We develop an speculative segmented sum algorithm by generating auxiliary information on-the-fly and eliminating costly pre- and post-processing on empty rows. 

  \item We evaluate our CSR-based SpMV algorithm on a widely-adopted benchmark suite and achieve stable SpMV performance independent of the sparsity structure of input matrix. 
  

\end{itemize}

On a benchmark suite composed of 20 matrices with diverse sparsity structures, our approach greatly outperforms the row block methods for the CSR-based SpMV running on GPU cores of heterogeneous processors. On an Intel heterogeneous processor, the experimental results show that our method obtains up to 6.90x and on average 2.57x speedup over an OpenCL implementation of the CSR-vector algorithm in CUSP running on its GPU cores. On an AMD heterogeneous processor, our approach delivers up to 16.07x (14.43x) and on average 5.61x (4.47x) speedup over the fastest single (double) precision CSR-based SpMV algorithm from PARALUTION and an OpenCL version of CUSP running on its GPU cores. On an nVidia heterogeneous processor, our approach delivers up to 5.91x (6.20x) and on average 2.69x (2.53x) speedup over the fastest single (double) precision CSR-based SpMV algorithm from cuSPARSE and CUSP running on its GPU cores. 


The paper is organized as follows. We first introduce background knowledge about the CSR format, the CSR-based SpMV algorithms and heterogeneous processors in Section 2. Then we describe our CSR-based SpMV algorithm in Section 3. Moreover, we give and analyze our experimental results in Section 4. We review the related approaches in Section 5. 

\section{Background and Motivations}

\subsection{CSR Format and CSR-based SpMV algorithms}

The CSR format of a sparse matrix consists of three separate arrays: (1) row pointer array of size $m+1$, where $m$ is the number of rows of the matrix, (2) column index array of size $nnz$, where $nnz$ is the number of nonzero entries of the matrix, and (3) value array of size $nnz$. Hence the overall space complexity of the CSR format is $O(m+nnz)$. Below we show a sparse matrix $A$ of size $6\times 6$ and its CSR representation.

\[
A = \begin{bmatrix}
  a & 0 & b & 0 & 0 & c \\
  d & e & f & 0 & 0 & 0 \\
  0 & 0 & g & 0 & h & 0 \\
  0 & 0 & 0 & 0 & 0 & 0 \\
  0 & 0 & 0 & 0 & i & 0 \\
  0 & 0 & j  & k  & l  & 0
 \end{bmatrix}
\]

\begin{equation} \label{eq1}
\begin{split}
row\; pointer & = \begin{bmatrix} 0, 3, 6, 8, 8, 9, 12\end{bmatrix} \\
column\; index & = \begin{bmatrix}0, 2, 5,\;\; 0, 1, 2,\;\; 2, 4,\;\; 4,\;\; 2, 3, 4\end{bmatrix} \\
value & = \begin{bmatrix}a, b, c,\;\; d, e, f,\;\; g, h,\;\;  i,\;\;\; j, k, l\end{bmatrix}. \nonumber
\end{split}
\end{equation}

Assume we have an input dense vector
\begin{equation} \label{eq1}
\begin{split}
x^T & = \begin{bmatrix}1, 2, 3, 4, 5, 6\end{bmatrix},\nonumber
\end{split}
\end{equation}
we can obtain a dense vector $y$ by multiplying the sparse matrix $A$ by the vector $x$:
\begin{equation} \label{eq1}
\begin{split}
y^T & = \begin{bmatrix}a+3b+6c, \;\;d+2e+3f, \;\;3g+5h, \;\;0, \;\;5i, \;\;3j+4k+5l\end{bmatrix}.  \nonumber
\end{split}
\end{equation}

The straightforward way to multiply $A$ with $x$ on a multicore processor is assigning a row block (i.e., multiple rows) to each core. Since the row pointer array records offset information of column index and value of nonzero entries, each core can easily position data in $A$ and $x$. Then generating corresponding entries of $y$ merely needs some simple multiply and add operations. For example, we assume that using a six-core processor for the above SpMV operation and each core is responsible for one row. We can notice that the cores calculating the first, the second and the sixth rows are busier than the other cores. Meanwhile, the core doing the fourth row is actually idle while the other cores are working. Therefore, the row block method cannot naturally handle load balance on multicore processors. On co-processors composed of a large amount of lightweight single instruction, multiple data (SIMD) units, the problem can heavily degrade performance of SpMV operation. Even though many strategies, such as vectorization~\cite{Bell:Implementing, Su:clSpMV, Williams:Optimization}, data streaming~\cite{Greathouse:Efficient}, memory coalescing~\cite{Deng:Taming}, static or dynamic binning~\cite{Greathouse:Efficient, Ashari:Fast}, Dynamic Parallelism~\cite{Ashari:Fast} and dynamic row distribution~\cite{Liu:LightSpMV}, have been proposed for the row block method, it is still impossible to achieve nearly perfect load balancing in general sense, simply since row sizes are irregular and unpredictable. 

The other method of computing the CSR-based SpMV is utilizing a segmented sum algorithm. This method first generates a segment descriptor of size $nnz$. The descriptor marks the first nonzero entry of each non-empty row as 1 (or equivalently \texttt{TRUE}) and the other nonzero entries as 0 (or equivalently \texttt{FALSE}). Using the above 6-by-6 sparse matrix as an example, we have 
\begin{equation} \label{eq1}
\begin{split}
segment\; descriptor & = [1, 0, 0,\;\; 1, 0, 0,\;\; 1, 0,\;\; 1,\;\; 1, 0, 0]. \nonumber
\end{split}
\end{equation}
Then an element-wise product array of size $nnz$ is allocated and filled by calculating 
\begin{equation} \label{eq1}
\begin{split}
product[i] = x[column\; index[i]]\times value[i], i\in[0,nnz). \nonumber
\end{split}
\end{equation}
The third step conducts a segmented sum operation on the product array by using segment information stored in the segment descriptor. Finally, the sum in each segment is stored to a contiguous location in $y$. 

We can see that the segmented sum method can achieve nearly perfect load balance in the nonzero entry space. However, this method has two obvious drawbacks: (1) since the segment descriptor is binary, this method is unable to recognize empty rows, thus a pre-processing (squeezing out possible empty rows) is required for calculating a ``clean'' row pointer array, and a post-processing (adding zeros to proper locations) is needed for a correct resulting vector $y$, and (2) this method requires more expensive global synchronizations and global memory access than the row block method (which needs only one single kernel launch). Therefore, in practice, the segmented sum method is not necessarily faster than the row block methods.

\subsection{Heterogeneous Processors}

Compared to homogeneous chip multiprocessors such as CPUs and GPUs, the heterogeneous processors are able to combine different types of cores into one chip. Thus heterogeneous processors offer more flexibilities in architecture design space. Because of mature CPU and GPU architectures and applications, CPU-GPU integrated heterogeneous processor with multiple instruction set architectures (ISAs) is the most widely adopted choice. Representatives of this model include AMD Accelerated Processing Units (APUs)~\cite{Branover:Llano, AMD:Compute}, Intel multi-CPU and GPU system-on-a-chip (SoC) devices~\cite{Damaraju:22nm}, nVidia Echelon heterogeneous GPU architecture~\cite{Keckler:GPUs}, and many mobile processors (e.g., nVidia Tegra~\cite{nVidia:Tegra} and Qualcomm Snapdragon~\cite{Qualcomm:Snapdragon}).

\begin{figure}[h!t]
\centering
\subfloat[Heterogeneous processor with separate last level cache]{\epsfig{file=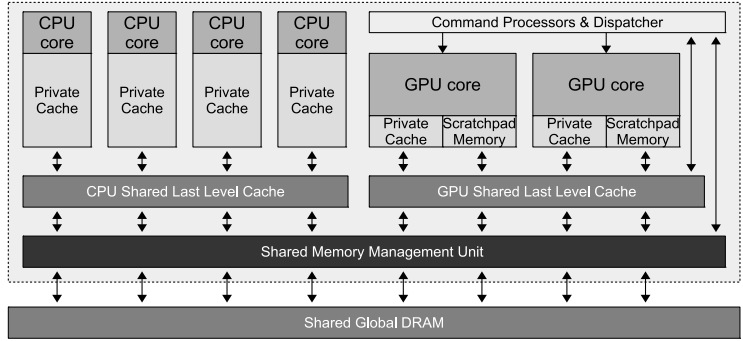, trim=0in 3.6in 0in 0in, width=3.8in}}
\vspace{3mm}
\qquad
\subfloat[Heterogeneous processor with CPU-GPU shared last level cache]{\epsfig{file=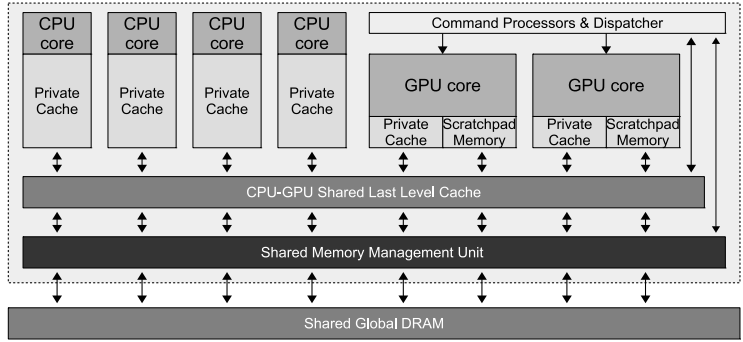, trim=0in 3.6in 0in 0in, width=3.8in}}
\caption{Block diagrams of two representative heterogeneous processors.}
\label{spmv.parco.fig.hcmps}
\end{figure}

Figure~\ref{spmv.parco.fig.hcmps} shows two block diagrams of heterogeneous processors used as experimental testbed in this paper. In general, a heterogeneous processor consists of four major parts: (1) a group of CPU cores with hardware-controlled caches, (2) a group of GPU cores with shared command processors, software-controlled scratchpad memory and hardware-controlled caches, (3) shared memory management unit, and (4) shared global dynamic random-access memory (DRAM). The last level cache of the two types of cores can be separate as shown in Figure~\ref{spmv.parco.fig.hcmps}(a) or shared as shown in Figure~\ref{spmv.parco.fig.hcmps}(b). 

The CPU cores have higher single-thread performance due to out-of-order execution, branch prediction and large amounts of caches. The GPU cores execute massively parallel lightweight threads on SIMD units for higher aggregate throughput. The two types of compute units have completely different ISAs and separate cache sub-systems. 

In this paper, our experiments run on three different platforms (shown in Table~\ref{spmv.parco.tab.testbed}), the platforms from AMD and nVidia are based on the design of Figure~\ref{spmv.parco.fig.hcmps}(a); the Intel platform uses the design of Figure~\ref{spmv.parco.fig.hcmps}(b). Note that in the current AMD APU architecture, although the two types of cores have separate last level caches, the GPU cores are able to snoop the last level cache on the CPU side. 


Compared to loosely-coupled CPU-GPU heterogeneous systems, the two types of cores in a heterogeneous processor share one single unified address space instead of using separate address spaces (i.e., system memory space and GPU device memory space). One obvious benefit is avoiding data transfer through connection interfaces (e.g., PCIe link), which is one of the most well known bottlenecks of co-processor computing~\cite{Gregg:Where}. Additionally, GPU cores can access more memory by paging memory to and from disk. Further, the consistent pageable shared virtual memory can be fully or partially coherent, meaning that much more efficient CPU-GPU interactions are possible due to eliminated heavyweight synchronization (i.e., flushing and GPU cache invalidation). Currently, programming on the unified address space and low-overhead kernel launch are supported by HSA~\cite{HSA:Manual}, OpenCL~\cite{Munshi:The} and CUDA~\cite{Negrut:Unified}. 

\section{New Sparse Matrix-Vector Multiplication Algorithm}

\subsection{Data Decomposition}

We first evenly decompose nonzero entries of the input matrix to multiple small tiles for load balanced data parallelism. Here we define a tile as a 2D array of size $W\times T$. The width $T$ is the size of a thread-bunch, which is the minimum SIMD execution unit in a given vector processor. It is also known as wavefront in AMD GPUs or warp in nVidia GPUs. The height $W$ is the workload (i.e., the number of nonzero entries to be processed) of a thread. A tile is a basic work unit in matrix-based segmented sum method~\cite{Sengupta:Scan, Dotsenko:Fast}, which is used as a building block in our SpMV algorithm. Actually, the term ``tile'' is equivalent to the term ``matrix'' used in original description of the segmented scan algorithms~\cite{Sengupta:Scan, Dotsenko:Fast}. Here we use ``tile'' to avoid confusion between a work unit of matrix shape and a sparse matrix in SpMV. 

Since a thread-bunch can be relatively too small (e.g., as low as 8 in current Intel GPUs) to amortize scheduling cost, we combine multiple thread-bunches into one thread-group (i.e., work-group in OpenCL terminology or thread block in CUDA terminology) for possibly higher throughput. We define $B$ to denote the number of thread-bunches in one thread-group. Additionally, we let each thread-bunch compute $S$ contiguous tiles. Thus higher on-chip resource reuse and faster global synchronization are expected.

\begin{figure}[ht]
\centering
\epsfig{file=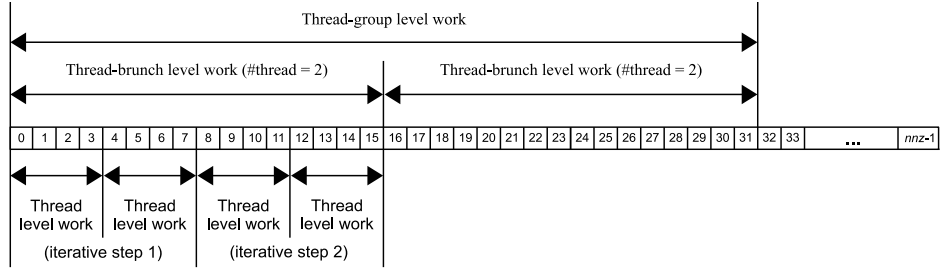, trim=0in 5in 0.5in 0.1in, width=5in}
\caption{Data decomposition on the nonzero entry space. $nnz$ nonzero entries are assigned to multiple thread-groups. In this case, each thread-group consists of 2 thread-bunches (i.e., $B=2$). The number of threads in each thread-bunch is equal to 2 (i.e., $T=2$). The workload per thread is 4 (i.e., $W=4$). The number of iterative steps in each thread-bunch is 2 (i.e., $S=2$).}
\label{spmv.parco.fig.datadecomposation}
\end{figure}

Therefore, we can calculate that each thread-group deals with $BSWT$ nonzero entries. Thus the whole nonzero entry space of size $nnz$ can be evenly assigned to $\lceil nnz / (BSWT)\rceil$ thread-groups. Figure~\ref{spmv.parco.fig.datadecomposation} shows an example of the data decomposition. In this example, we set $B=2$, $S=2$, $W=4$, and $T=2$. Thus each thread-group is responsible for 32 nonzero entries. Then $\lceil nnz / 32\rceil$ thread-groups are dispatched. 


\subsection{Algorithm Description}

Our CSR-based SpMV is based on fundamental segmented sum algorithm, which guarantees load balanced computation in the nonzero entry space. While utilizing segmented sum as a building block in our SpMV algorithm, we have three main performance considerations: (1) the segment descriptor needs to be generated in on-chip memory at runtime to reduce overhead of global memory access, (2) empty rows must be recognized and processed without calling specific pre- and post-processing functions, and (3) taking advantages of both types of cores in a heterogeneous processor. Hence we improve the basic segmented sum method to meet the above performance requirements.

The algorithm framework includes two main stages: (1) speculative execution stage, and (2) checking prediction stage. The first stage speculatively executes SpMV operation and generates a possibly incorrect resulting vector $y$. Here the term ``incorrect'' means that the layout of entries in $y$ can be incorrect, but the entries are guaranteed to be numerically identified. Then in the second stage we check whether or not the speculative execution is successful. If the prediction is wrong, a data re-arrangement will be launched for getting a completely correct $y$.

\begin{figure}[!t]
\centering
\epsfig{file=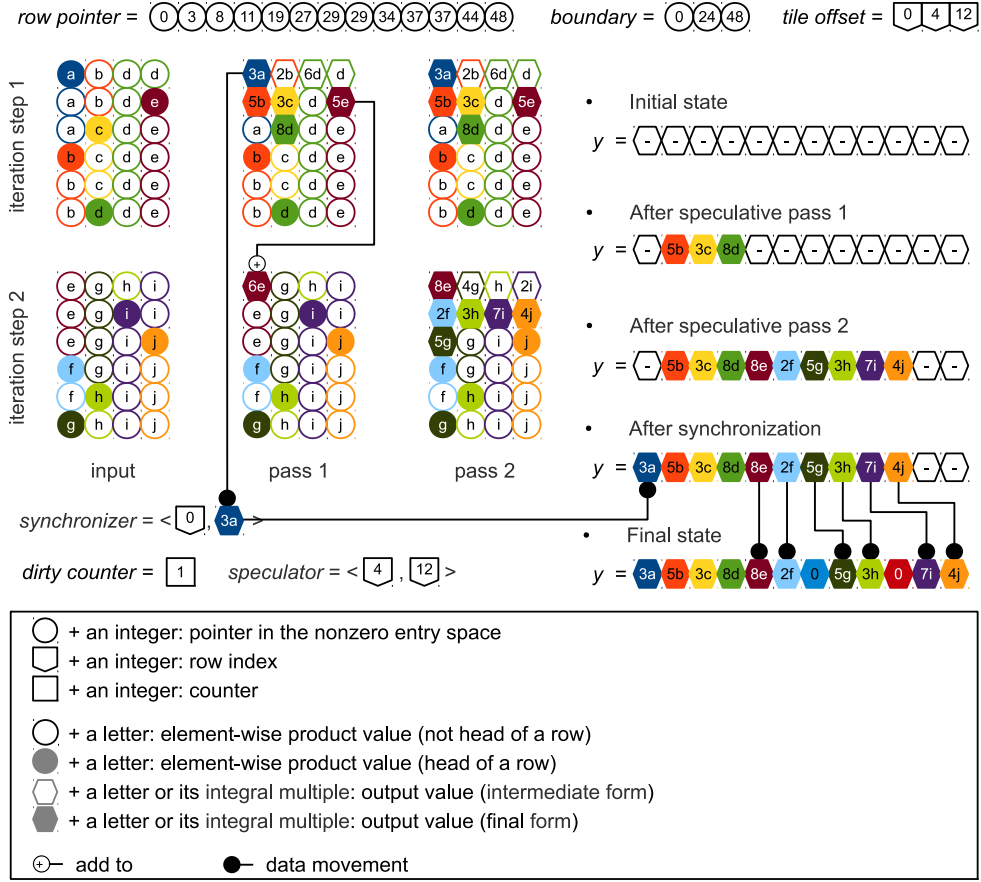, trim=0.1in 2.1in 1.5in 0in, width=5.2in}
\caption{An example of our CSR-based SpMV algorithm. The input sparse matrix contains 48 nonzero entries in 12 rows (10 non-empty rows and 2 empty rows). One thread-bunch composed of 4 threads is launched in this 2-iteration process. The arrays \textit{synchronizer} and \textit{speculator} store tuples (shown with angular brackets).}
\label{spmv.parco.fig.example}
\end{figure}

We first give an example of our algorithm and use it in the following algorithm description. Figure~\ref{spmv.parco.fig.example} plots this example. The input sparse matrix includes 12 rows (2 of them are empty) and 48 nonzero entries. We set $B$ to 1, $S$ to 2, $T$ to 4 and $W$ to 6. This setting means that one thread-group is composed of one thread-bunch of size 4; each thread-bunch runs 2 iteration steps. Before GPU kernel launch, three containers, \textit{synchronizer}, \textit{dirty\_counter} and \textit{speculator}, are pre-allocated in DRAM for global synchronization and speculative execution. Algorithm~\ref{spmv.parco.alg.spmv} in Appendix A lists pseudo code of the first stage.

The \textbf{speculative execution stage} includes the following steps: (1) positioning a range of row indices for nonzero entries in a given tile, (2) calculating segment descriptor based on the range, (3) conducting segmented sum on the tile, (4) saving partial sums to the computed index range in vector $y$. This stage also has some input-triggered operations such as labeling a tile with empty rows.

First, each thread-bunch executes binary search of $S+1$ tile boundaries on the CSR row pointer array. Then we obtain corresponding row indices and store them in a scratchpad array \textit{tile offset} of size $S+1$. The results of the binary search are starting and ending row indices of the nonzero entries in each tile. Thus each tile knows the locations to store generated partial sums. Lines 3--7 of Algorithm~\ref{spmv.parco.alg.spmv} give a code expression of this step. In our example shown in Figure~\ref{spmv.parco.fig.example}, the 2 tiles of size 24 have 3 boundaries \{0, 24, 48\}. The results of binary search of \{0, 24, 48\} on the CSR row pointer array are \{0, 4, 12\}. Note that the binary search needs to return the rightmost match, if multiple slots have the same value.

Then each thread-bunch executes an iteration of $S$ steps. Lines 8--59 of Algorithm~\ref{spmv.parco.alg.spmv} give code expression of this step. Each iteration deals with one tile. By calculating offset between the left boundary of a tile and the covered row indices, a local segment descriptor is generated (lines 14--21 in Algorithm~\ref{spmv.parco.alg.spmv}). For example, the left boundary of the second tile is 24 and its row index range is 4--12. We need to compute offset between 24 and the row pointer \{19, 27, 29, 29, 34, 37, 37, 44, 48\}. Then we obtain a group of offsets \{-5, 3, 5, 5, 10, 13, 13, 20, 24\}. After removing duplicate values and overflowed values on the left and the right sides, the effective part \{3, 5, 10, 13, 20\} in fact implies local segment descriptor for the current tile. We can easily convert it to a binary expression \{0, 0, 0, 1, 0, 1, 0, ... , 0, 0, 1, 0, 0, 0\} through a scatter operation in on-chip scratchpad memory. Moreover, since each tile is an independent work unit, the first bit of its segment descriptor should be \texttt{TRUE}. Thus the final expression becomes \{1, 0, 0, 1, 0, 1, 0, ... , 0, 0, 1, 0, 0, 0\}. In Figure~\ref{spmv.parco.fig.example}, the filled and empty circles are heads (i.e., 1s or \texttt{TRUE}s) and body (i.e., 0s or \texttt{FALSE}s) of segments, respectively.



While generating the segment descriptor, each thread detects whether or not its right neighbor wants to write to the same slot. If yes (like the duplicate offset information \{..., 5, 5, ...\} and \{..., 13, 13, ...\} in the above example), we can make sure that this tile contains at least one empty row, since an empty row is expressed as two contiguous indices of the same value in the CSR row pointer array. Then we mark this tile as ``dirty'' (line 19 in Algorithm~\ref{spmv.parco.alg.spmv}). Further, the \textit{dirty counter} array stored in DRAM is incremented by atomic operation, and this tile's offset is recorded in the \textit{speculator} array (lines 53--58 in Algorithm~\ref{spmv.parco.alg.spmv}). In our example, \textit{dirty counter} is 1 and \textit{speculator} array has a pair of offsets \{$\langle$4, 12$\rangle$\} ((shown with angular brackets in Figure~\ref{spmv.parco.fig.example}).

Then we calculate and save element-wise products in scratchpad memory, based on its nonzero entries' column indices, values and corresponding values in the vector $x$. Lines 22--26 of Algorithm~\ref{spmv.parco.alg.spmv} show code expression of this step. When finished, we transmit the sum of the last segment to an intermediate space for the next iteration (lines 27--31 in Algorithm~\ref{spmv.parco.alg.spmv}). In our example, the first tile's last value $5e$ is transmitted to the next tile. Then we execute the matrix-based segmented sum (lines 32--33) on the tile. Because the segmented sum algorithm used here is very similar to the method described in~\cite{Blelloch:Segmented}, we refer the reader to~\cite{Blelloch:Segmented} and several pervious GPU segmented sum algorithms~\cite{Sengupta:Scan, Dotsenko:Fast} for details. But note that compared to~\cite{Blelloch:Segmented}, our method makes one difference: we store partial sums in a compact pattern (i.e., values are arranged in order from the first location in the thread work space), but not save them to locations of corresponding segment heads. For this reason, we need to record the starting position and the number of partial sums. Then we can use an ordinary exclusive scan operation (lines 34--35) for obtaining contiguous indices of the partials sums in $y$. In Figure~\ref{spmv.parco.fig.example}, we can see that the partial sums (expressed as filled hexagons) are aggregated in the compact fashion. Note that empty hexagons are intermediate partial sums, which are already added to the correct position of segment heads.

Finally, we store the partial sums to known locations in the resulting vector. Lines 36--52 of Algorithm~\ref{spmv.parco.alg.spmv} show code expression. As an exception, the sum result of the first segment in a thread-bunch is stored to the \textit{synchronizer} array (lines 40--43), since the first row of each thread-bunch may cross multiple thread-bunch. This is a well known issue while conducting basic primitives, such as reduction and prefix-sum scan, using more than one thread-group that cannot communicate with each other. In fact, atomic add operation can be utilized to avoid the global synchronization. But we choose not to use relatively slow global atomic operations and let a CPU core to later on finish the global synchronization. Lines 62--68 of Algorithm~\ref{spmv.parco.alg.spmv} show the corresponding code expression. Since the problem size (i.e., $\lceil nnz / (SWT)\rceil$) can be too small to saturate a GPU core, a CPU core is in fact faster for accessing short  arrays linearly stored in DRAM. Taking the first tile in Figure~\ref{spmv.parco.fig.example} as an example, its first partial sum is $3a$, which is stored with its global index 0 to the \textit{synchronizer}. After that, the value $3a$ is added to position 0 of $y$.

When the above steps are complete, the resulting vector is numerically identified, except that some values generated by dirty tiles are not in their correct locations. In Figure~\ref{spmv.parco.fig.example}, we can see that after synchronization, vector $y$ is already numerically identified to its final form, but entries $5g$, $3h$, $7i$ and $4j$ generated by the second tile are located in wrong slots.

The \textbf{checking prediction stage} first checks value of the \textit{dirty counter} array. If it is zero, the previous prediction is correct and the result of the first stage is the final result; if it is not zero, the predicted entries generated by dirty tiles are scattered to their correct positions in the resulting vector. In this procedure, the CSR row pointer array is required to be read for getting correct row distribution information. Again, we use a CPU core for the irregular linear memory access, which is more suitable for cache sub-systems in CPUs. In our example, entries $5g$, $3h$, $7i$ and $4j$ are moved to their correct positions. Then the SpMV operation is done.

\subsection{Complexity Analysis}

Our CSR-based SpMV algorithm pre-allocates three auxiliary arrays, \textit{synchronizer}, \textit{dirty counter} and \textit{speculator}, in DRAM. The space complexity of \textit{synchronizer} is $\lceil nnz / (SWT)\rceil$, equivalent to the number of thread-bunches. The size of \textit{dirty counter} is constant 1. The \textit{speculator} array needs a size of $\lceil nnz / (WT)\rceil$, equivalent to the number of tiles. Since $W$ and $T$ are typically set to relatively large values, the auxiliary arrays merely slightly increase overall space requirement.

For each thread-bunch, we executes $S+1$ binary searches in the row pointer array of size $m+1$. Thus $O(\lceil nnz / (SWT) \rceil \times (S+1)\times \log_2 (m+1))=O(nnz \log_2 (m) / WT)$ is work complexity of this part. On the whole, generating segment descriptor needs $O(m)$ time. Collecting element-wise products needs $O(nnz)$ time. For each tile, segmented sum needs $O(WT+\log_2 (T))$ time. Thus all segmented sum operations need $O(\lceil nnz / (WT) \rceil (WT+\log_2 (T)))=O(nnz + nnz \log_2 (T) / WT)$ time. Saving entries to $y$ needs $O(m)$ time. Synchronization takes $O(\lceil nnz / (SWT)\rceil)=O(nnz/SWT)$ time. Possible re-arrangement needs $O(m)$ time in the worst case. Thus overall work complexity of our CSR-based SpMV algorithm is $O(m+nnz+nnz(\log_2 (m)+\log_2 (T))/WT)$.

\subsection{Implementation Details}

Based on the above analysis, we can see that when the input matrix is fixed, the cost of our SpMV algorithm only depends on two parameters: $T$ and $W$. In our algorithm implementation, $T$ is set to SIMD length of the used processor. Choosing $W$ needs to consider the capacity of on-chip scratchpad memory. The other two parameters $B$ and $S$ are empirically chosen. Table~\ref{spmv.parco.tab.parameters} shows the selected parameters. Note that double precision is not currently supported in Intel OpenCL implementation for its GPUs. 

\begin{table*}[!ht]
\small 
\caption{The selected parameters}
\label{spmv.parco.tab.parameters}
\centering
\begin{tabular}{  >{\centering}m{1.5cm}   >{\centering}m{1.25cm}   >{\centering}m{1.25cm}  >{\centering}m{1.25cm}  >{\centering}m{1.25cm}  >{\centering}m{1.25cm} } 
\hline

Processor 		& Intel	& \multicolumn{2}{c}{AMD} & \multicolumn{2}{c}{nVidia} \tabularnewline \hline
Precision 	&  32-bit single 	&  32-bit single  & 64-bit double  &  32-bit single & 64-bit double   \tabularnewline \hline
$T$ 	&  8 	&  64  & 64  & 32  & 32   \tabularnewline 
$W$ 	&   16	&  16  & 8  &  8 &  4  \tabularnewline 
$B$ 	&  4 	&  2  &  2 & 5  &  5  \tabularnewline 
$S$ 	&  6 	&   2 & 5  & 7  &  7  \tabularnewline 
\hline
\end{tabular}
\end{table*}

We implement the first stage of our algorithm in OpenCL for the Intel and AMD platforms (and CUDA for the nVidia platform) for GPU execution and the second stage in standard C language running on the CPU part. Since our algorithm needs CPU and GPU share some arrays, we allocate all arrays in Shared Virtual Memory supported by OpenCL for the best performance. On the nVidia platform, we use Unified Memory in CUDA SDK.

\section{Experimental Results}

\subsection{Experimental Environments}

We use three heterogeneous processors, Intel Core i3-5010U, AMD A10-7850K APU and nVidia Tegra K1, for evaluating SpMV algorithms. Table~\ref{spmv.parco.tab.testbed} shows specifications of the three processors. All of them are composed of multiple CPU cores and GPU cores. The two types of cores in the Intel heterogeneous processor share a 3 MB last level cache. In contrast, GPU cores in the AMD heterogeneous processor can snoop the L2 cache of size 4 MB on the CPU side. Unlike those, the cache systems of the CPU part and the GPU part in the nVidia Tegra processor are completely separate. Note that currently the Intel GPU can run OpenCL program only on Microsoft Windows operating system. Also note that we use kB, MB and GB to denote $2^{10}$, $2^{20}$ and $2^{30}$ bytes, respectively; and use GFlop to denote $10^9$ flops.

\begin{table*}[!ht]
\tiny 
\caption{The test environments used in our experiments}
\label{spmv.parco.tab.testbed}
\centering
\begin{tabular}{  m{2.5cm}   >{\centering}m{1.3cm}  >{\centering}m{1.3cm}  >{\centering}m{1.3cm}  >{\centering}m{1.3cm}  >{\centering}m{1.3cm}  >{\centering}m{1.3cm} } 
\hline
Processor 		& \multicolumn{2}{c}{Intel Core i3-5010U}	& \multicolumn{2}{c}{AMD A10-7850K APU} & \multicolumn{2}{c}{nVidia Tegra K1} \tabularnewline \hline
Core type 	&  x86 CPU 	& GPU  &x86 CPU &GPU &ARM CPU &GPU\tabularnewline 
Codename 	&  Broadwell  & HD 5500	& Steamroller &GCN & Cortex A15 &Kepler\tabularnewline 
Cores @ clock (GHz)		 	& 2 @ 2.1	& 3 @ 0.9 	& 4 @ 3.7  & 8 @ 0.72  & 4 @ 2.3  & 1 @ 0.85\tabularnewline 
SP flops/cycle 	& 2$\times$32			&  3$\times$128  & 4$\times$8 			& 8$\times$128 & 4$\times$8 			& 1$\times$384\tabularnewline
SP peak (GFlop/s) 	&  134.4			& 345.6 & 118.4 			& 737.3 & 73.6 			& 327.2\tabularnewline
DP flops/cycle 	& 2$\times$16		& 3$\times$32   & 4$\times$4 			& 8$\times$8 & 2$\times$2 			& 1$\times$16\tabularnewline
DP peak (GFlop/s) 	&  67.2			& 86.4 & 59.2 			& 46.1 & 18.4 			& 13.6\tabularnewline 
L1 data cache 	& 4$\times$32 kB 		& 3$\times$4 kB & 4$\times$16 kB 			& 8$\times$16 kB & 4$\times$32 kB 			& 1$\times$16 kB\tabularnewline
L2 cache  	& 4$\times$256 kB	&3$\times$24 kB	& 2$\times$2 MB & Unreleased & 2 MB & 128 kB\tabularnewline 
L3 cache  	& N/A	&384 kB	& N/A & N/A & N/A & N/A\tabularnewline 
Scratchpad 		& N/A 			& 3$\times$64 kB & N/A			& 8$\times$64 kB & N/A			& 1$\times$48 kB\tabularnewline

Shared last level cache 	& \multicolumn{2}{c}{3 MB} 	& \multicolumn{2}{c}{N/A}  & \multicolumn{2}{c}{N/A}\tabularnewline 


DRAM  			& \multicolumn{2}{c}{Dual-channel DDR3-1600} 	& \multicolumn{2}{c}{Dual-channel DDR3-1600} & \multicolumn{2}{c}{Single-channel DDR3L-1866} \tabularnewline
DRAM capacity 	& \multicolumn{2}{c}{8 GB}		 	& \multicolumn{2}{c}{8 GB} & \multicolumn{2}{c}{2 GB} \tabularnewline
DRAM bandwidth  	& \multicolumn{2}{c}{25.6 GB/s}		& \multicolumn{2}{c}{25.6 GB/s}  & \multicolumn{2}{c}{14.9 GB/s}\tabularnewline \hline

Operating system 	& \multicolumn{2}{c}{Microsoft Windows 64-bit} 	&\multicolumn{2}{c}{Ubuntu Linux 14.04 64-bit}  &\multicolumn{2}{c}{Ubuntu Linux 14.04 32-bit} \tabularnewline
GPU driver version 			& \multicolumn{2}{c}{15.36} 			& \multicolumn{2}{c}{14.501} & \multicolumn{2}{c}{r19.2} \tabularnewline
Compiler	 			& \multicolumn{2}{c}{icc 15.0.2} & \multicolumn{2}{c}{gcc 4.8.2} & \multicolumn{2}{c}{gcc 4.8.2, nvcc 6.0.1}  \tabularnewline
Toolkit version 			& \multicolumn{2}{c}{ OpenCL 2.0} 			& \multicolumn{2}{c}{OpenCL 2.0}  & \multicolumn{2}{c}{CUDA 6.0}  \tabularnewline
\hline
\end{tabular}
\end{table*}

\subsection{Benchmark Suite}

To evaluate our method, we choose 20 unstructured matrices from the University of Florida Sparse Matrix Collection~\cite{Davis:The}. Table~\ref{spmv.parco.tab.benchmark} lists main information of the evaluated sparse matrices. The first 14 matrices of the benchmark suite have been widely used in previous work~\cite{Bell:Implementing, Su:clSpMV, Yan:yaSpMV, Ashari:An, Liu:CSR5, Williams:Optimization}. The last 6 matrices are chosen as representatives of irregular matrices extracted from graph applications, such as circuit simulation and optimization problems. 

The first 10 matrices are relatively regular, due to short distance between the average value and the maximum value of $nnz$/row. The other matrices are relatively irregular. In this context, `regular' is used for a sparse matrix including rows of roughly the same size. In contrast, an `irregular matrix' can have some very long rows and many very short rows. For example, matrices generated from power-law graphs can have a few rows with $O(n)$ nonzero entries and many rows with $O(1)$ nonzero entries.


\begin{table}[!ht]
\tiny 
\renewcommand{\arraystretch}{1.3}
\caption{Overview of evaluated sparse matrices}
\label{spmv.parco.tab.benchmark}
\centering
\begin{tabular}{ c  c  c  c  c  c }
\hline
\textbf{Name} & 
Dense & 
Protein & 
FEM/Spheres & 
FEM/Cantilever & 
Wind Tunnel \\
\textbf{Plot} &
\begin{minipage}[c]{0.12\columnwidth} \centering \includegraphics[width=0.6in]{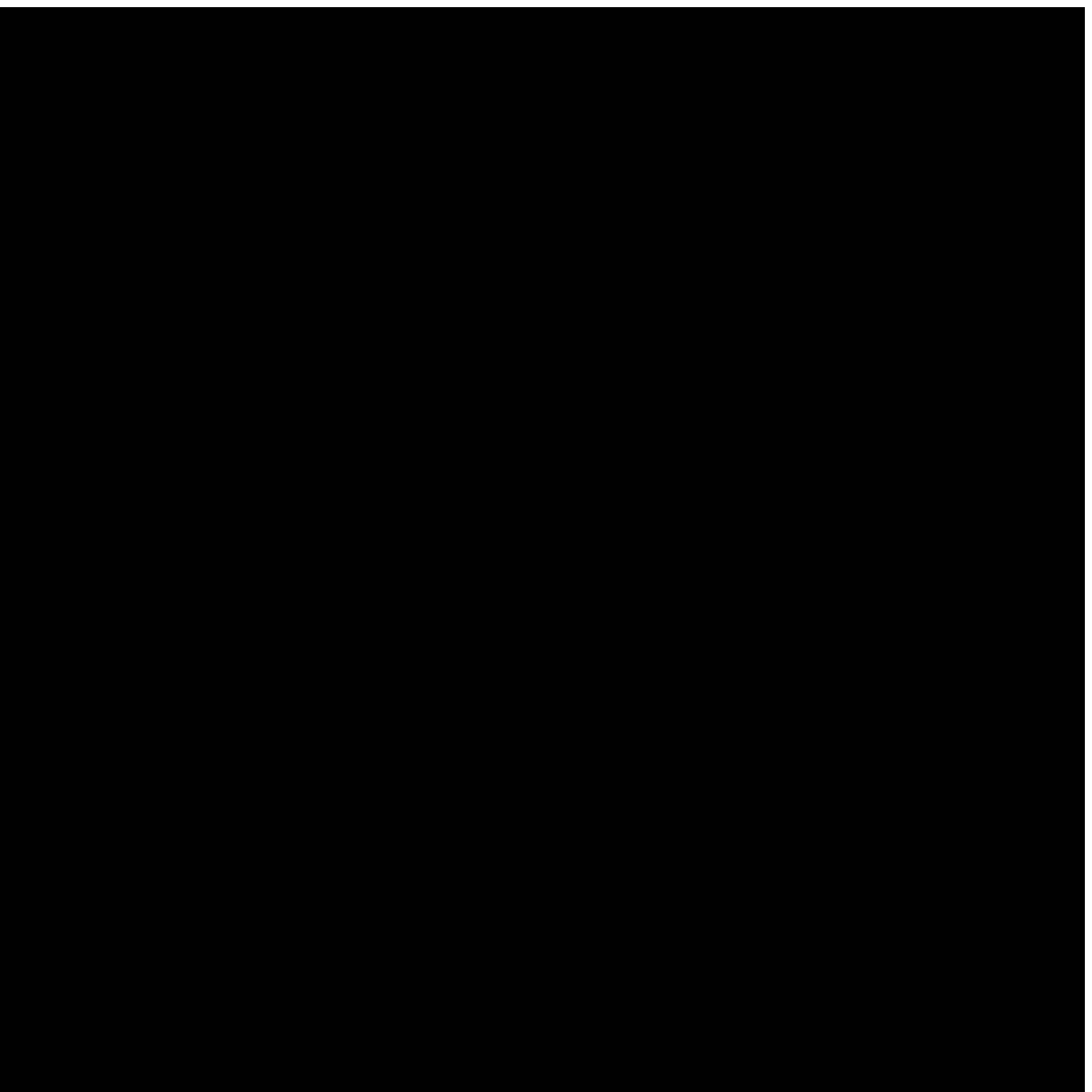} \end{minipage} &
\begin{minipage}[c]{0.12\columnwidth} \centering \includegraphics[width=0.6in]{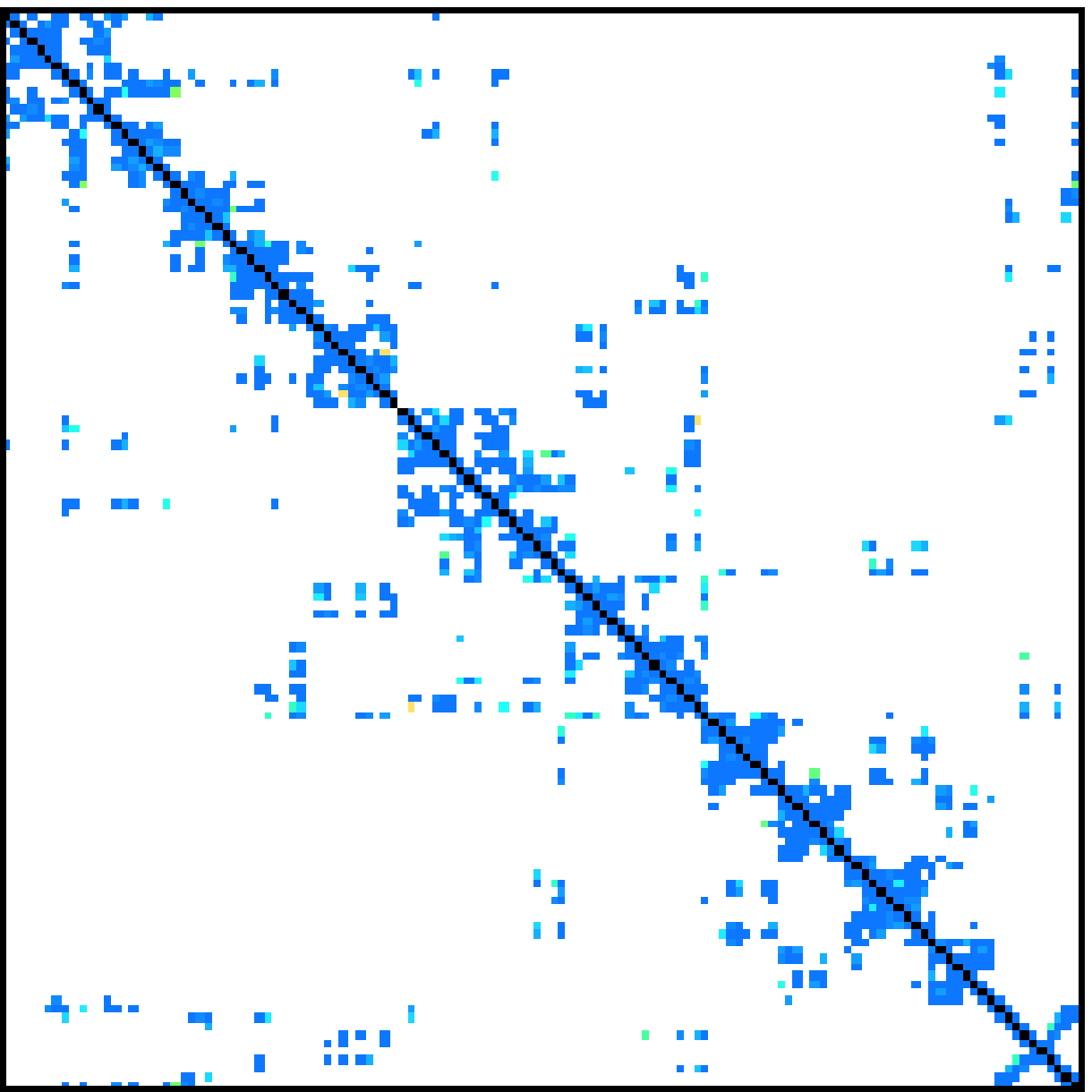} \end{minipage} &
\begin{minipage}[c]{0.12\columnwidth} \centering \includegraphics[width=0.6in]{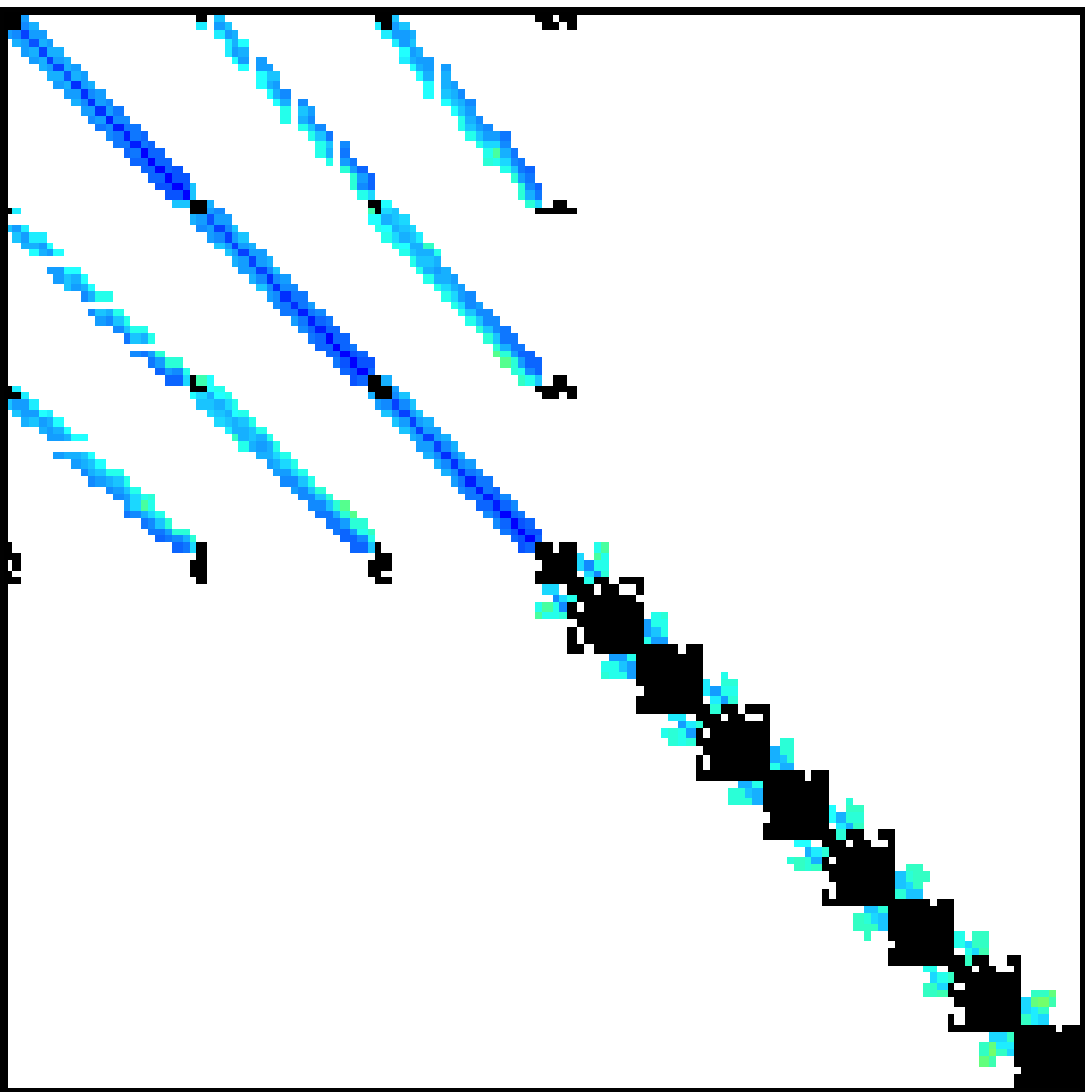} \end{minipage} &
\begin{minipage}[c]{0.12\columnwidth} \centering \includegraphics[width=0.6in]{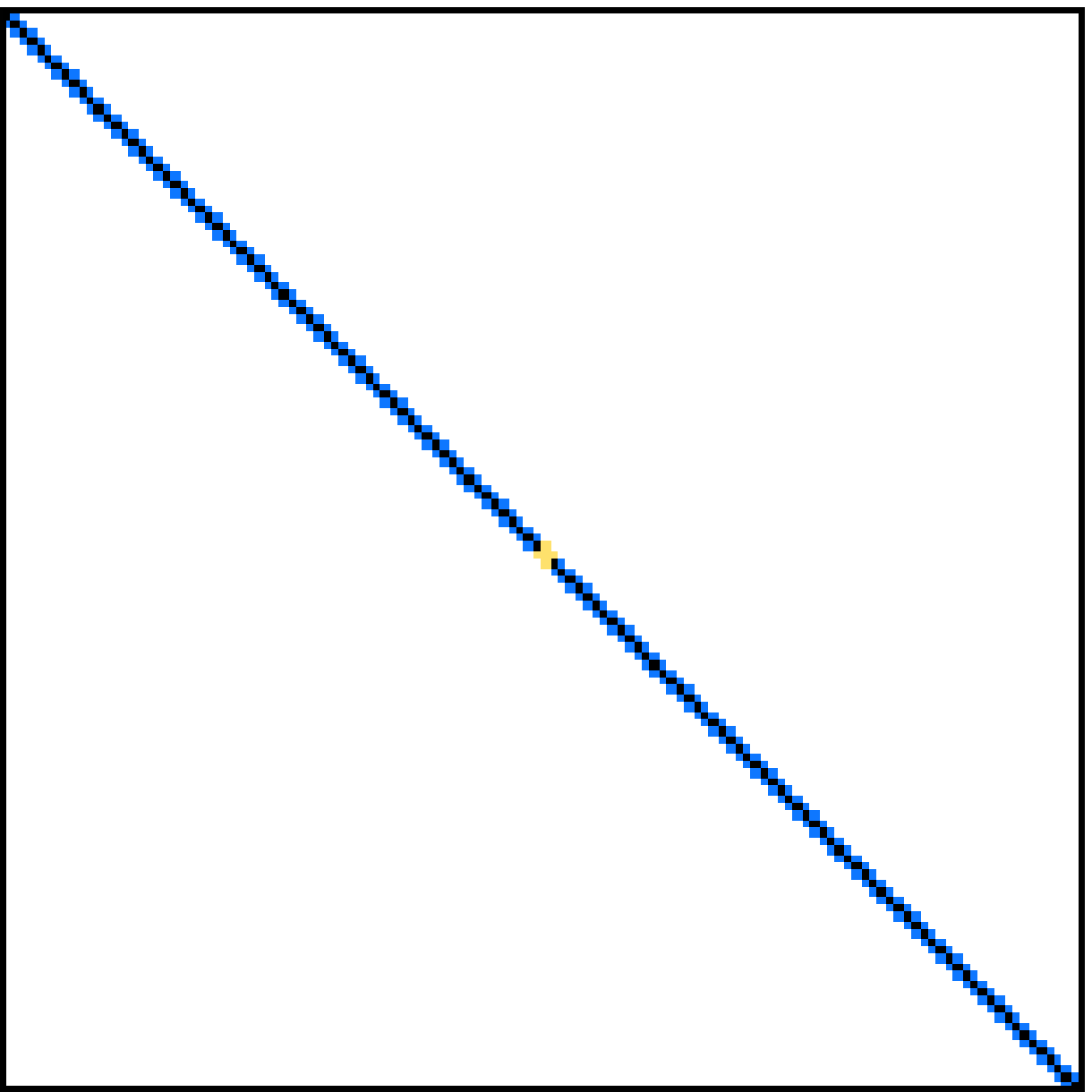} \end{minipage} &
\begin{minipage}[c]{0.12\columnwidth} \centering \includegraphics[width=0.6in]{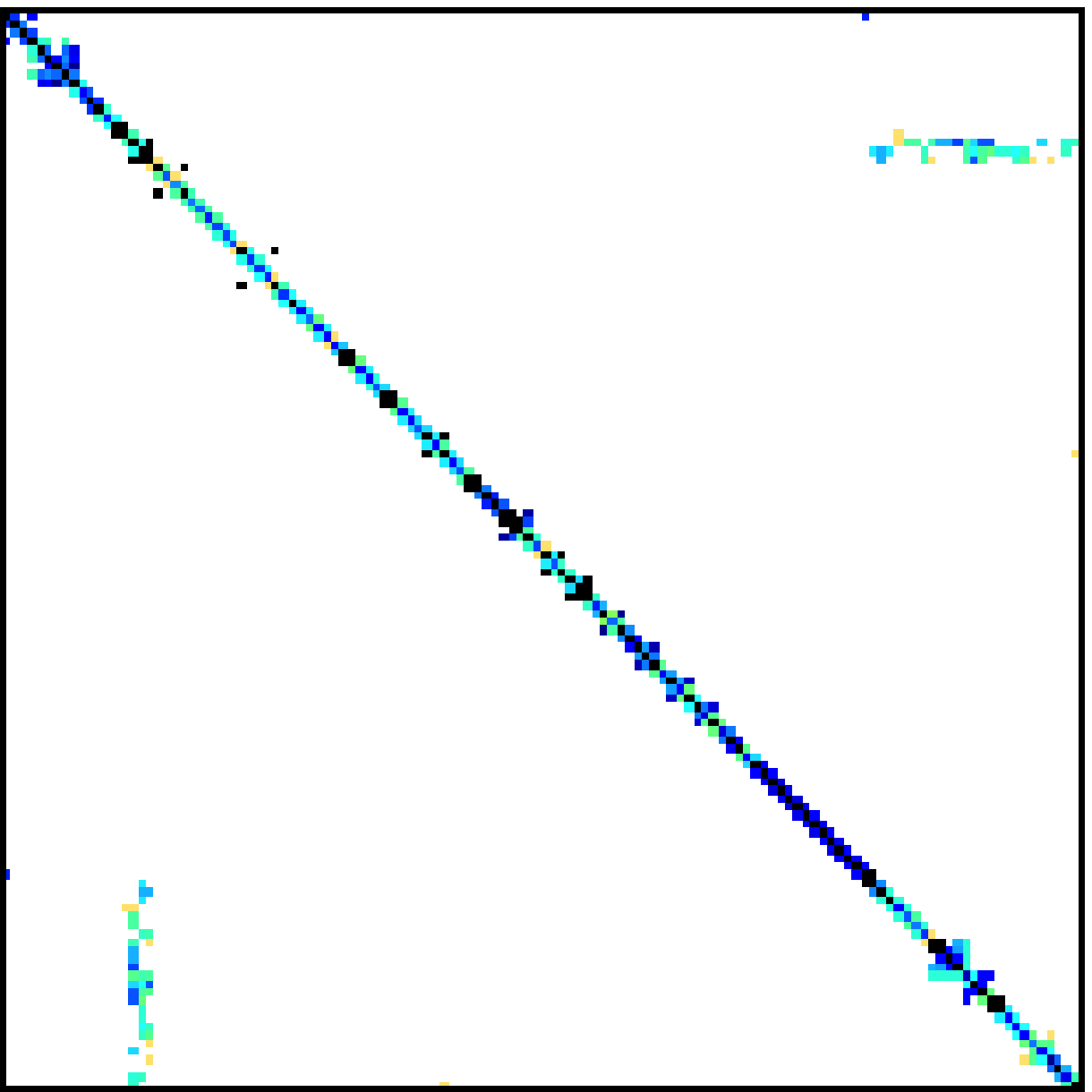} \end{minipage} \\ 
\textbf{Dimensions} & 
2 K $\times$ 2 K & 
36 K $\times$ 36 K & 
83 K $\times$ 83 K & 
62 K $\times$ 62 K & 
218 K $\times$ 218 K \\
\boldmath{$nnz$} & 
4.0 M & 
4.3 M & 
6.0 M & 
4.0 M & 
11.6 M \\
\begin{minipage}[c]{0.15\columnwidth} \centering \boldmath{$nnz/row$} \textbf{(min, avg, max)} \end{minipage}  & 
2 K, 2 K, 2 K & 
18, 119, 204 & 
1, 72, 81 & 
1, 64, 78 & 
2, 53, 180 \\
\hline

\textbf{Name} & 
FEM/Harbor & 
QCD & 
FEM/Ship  & 
Economics & 
Epidemiology \\
\textbf{Plot} &
\begin{minipage}[c]{0.12\columnwidth} \centering \includegraphics[width=0.6in]{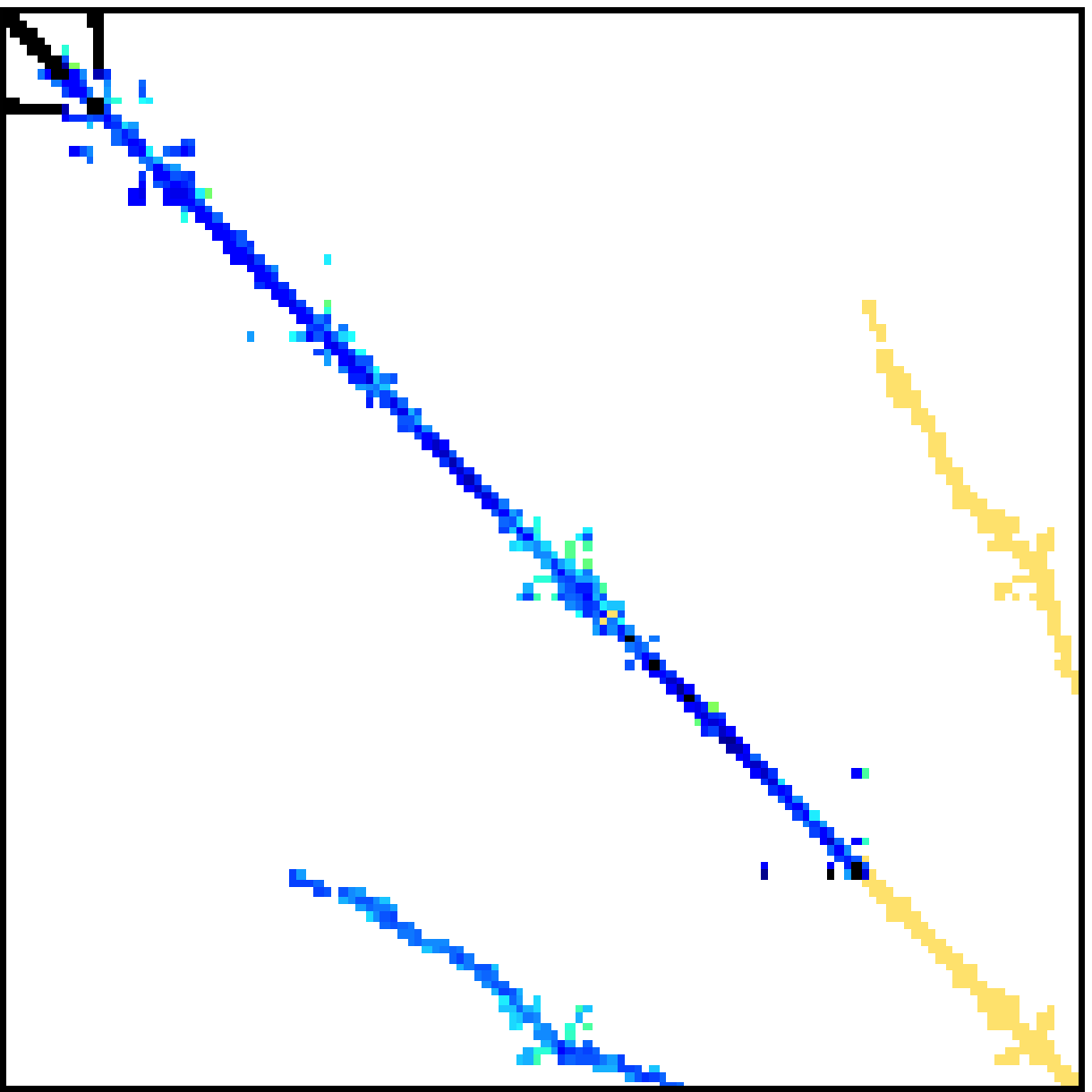} \end{minipage} &
\begin{minipage}[c]{0.12\columnwidth} \centering \includegraphics[width=0.6in]{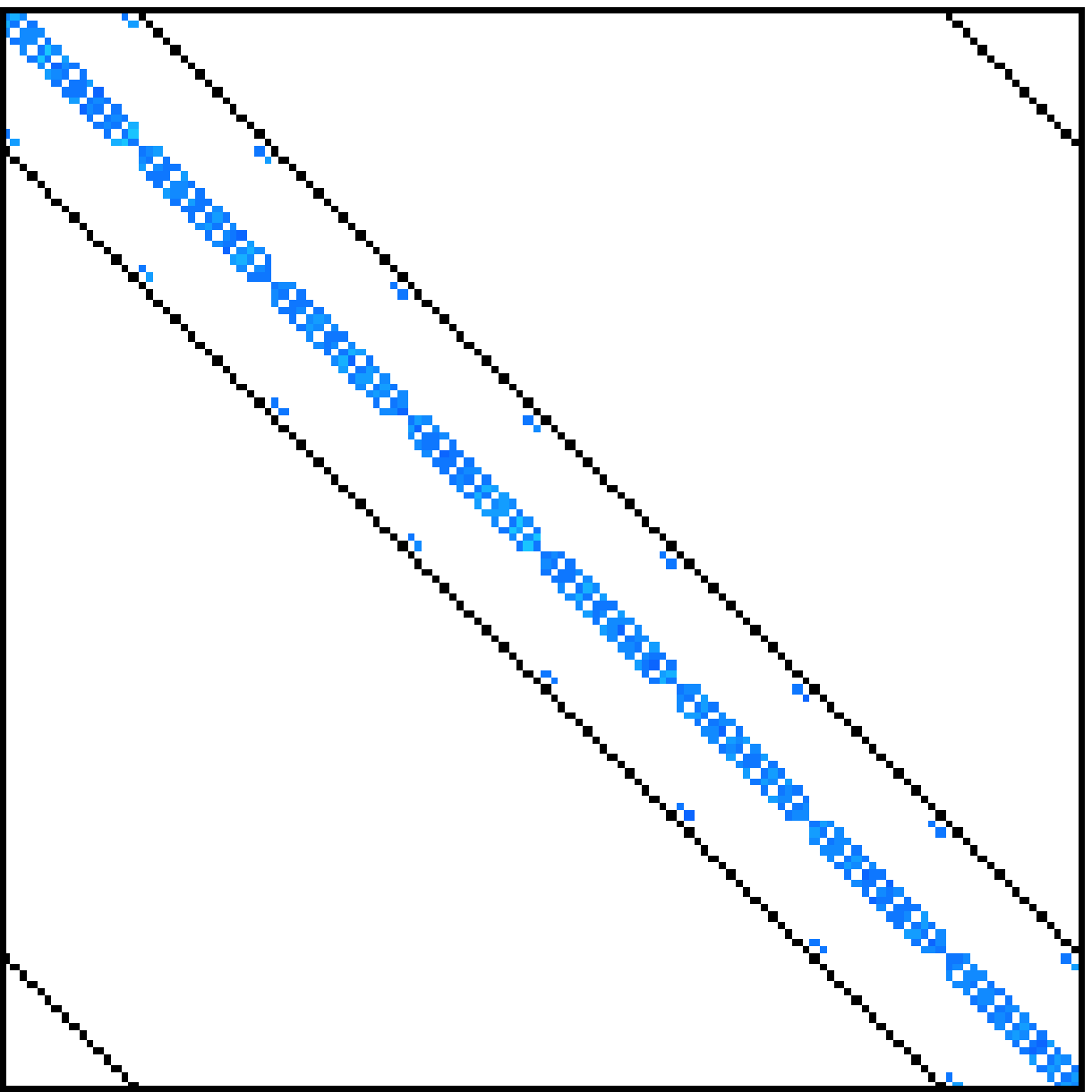} \end{minipage} &
\begin{minipage}[c]{0.12\columnwidth} \centering \includegraphics[width=0.6in]{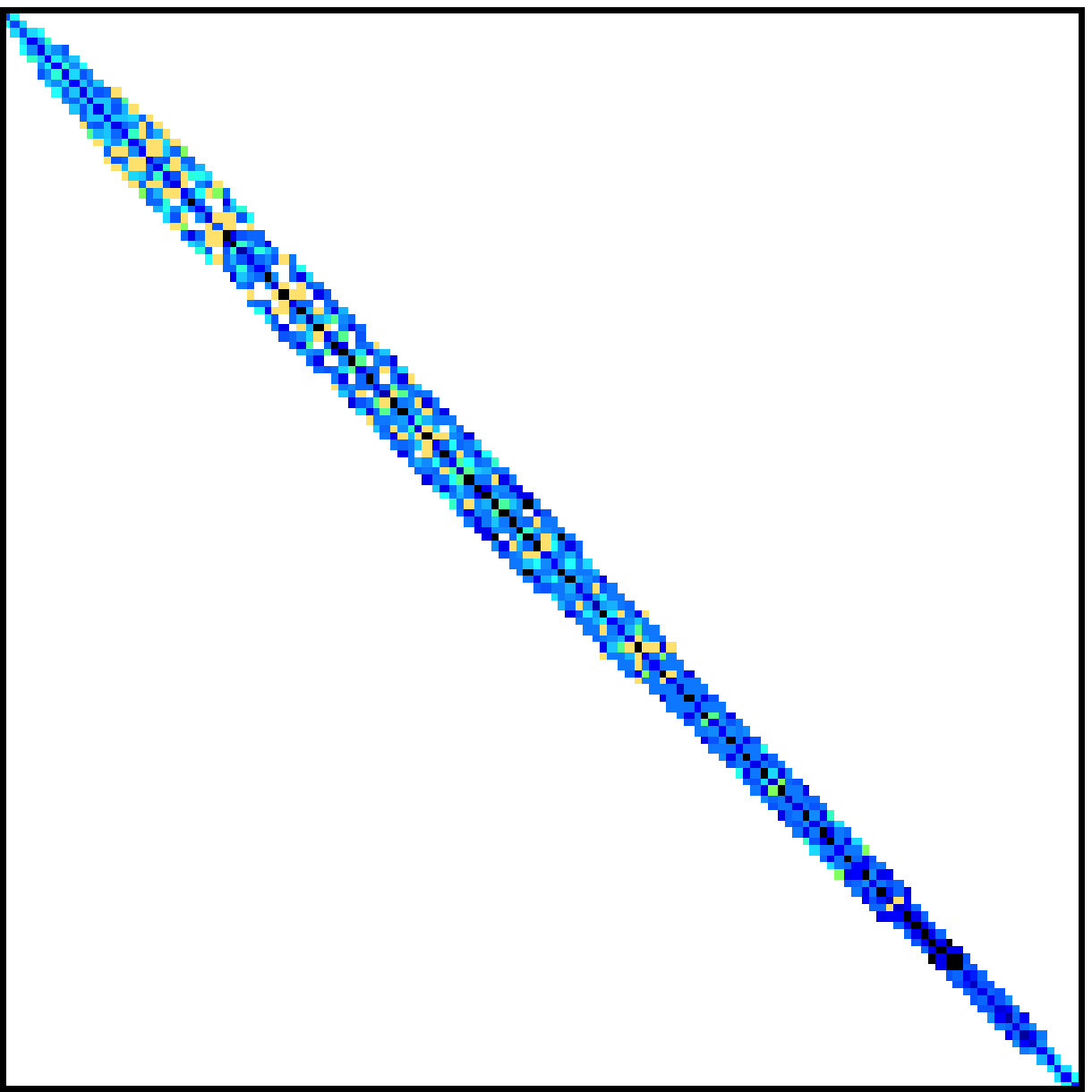} \end{minipage} &
\begin{minipage}[c]{0.12\columnwidth} \centering \includegraphics[width=0.6in]{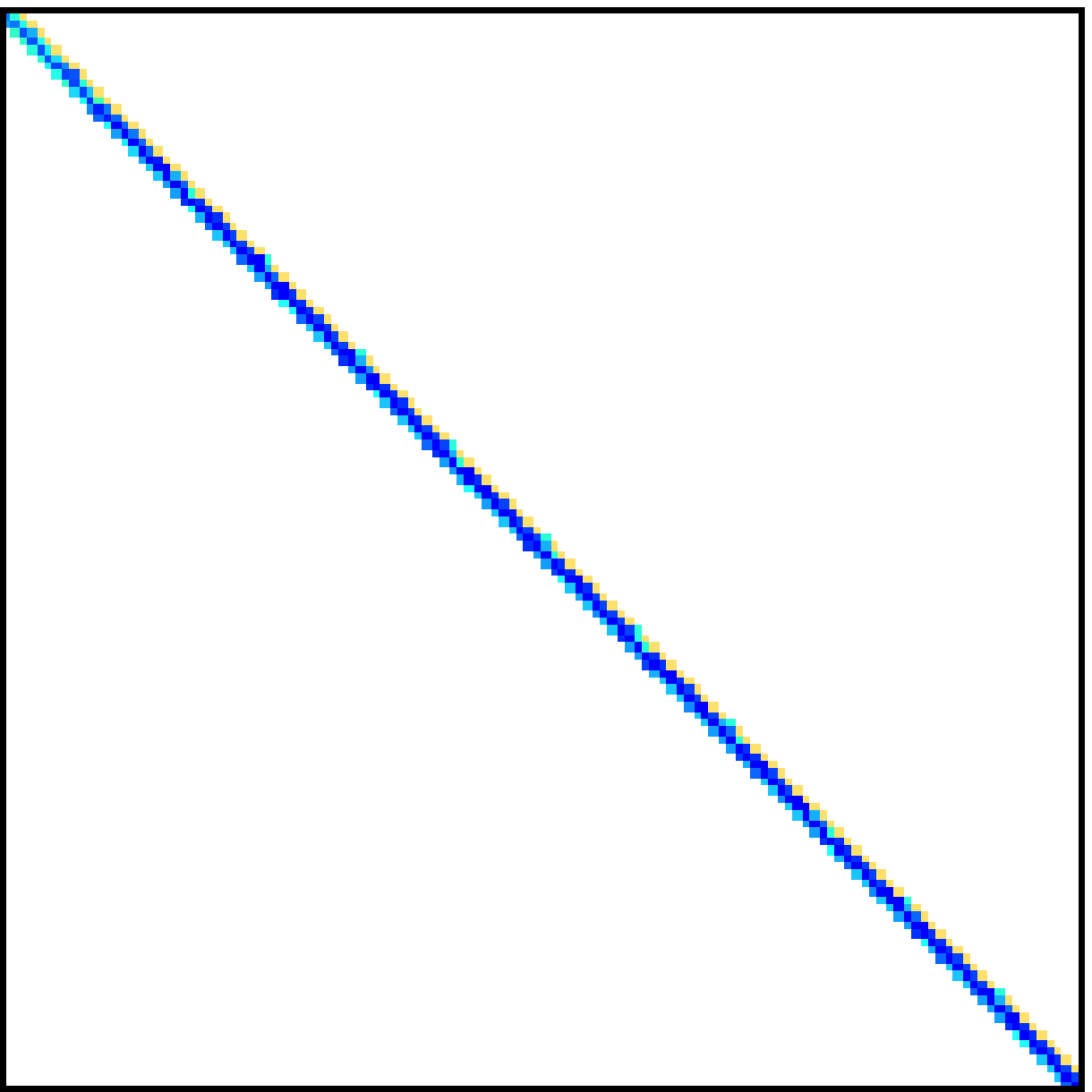} \end{minipage} &
\begin{minipage}[c]{0.12\columnwidth} \centering \includegraphics[width=0.6in]{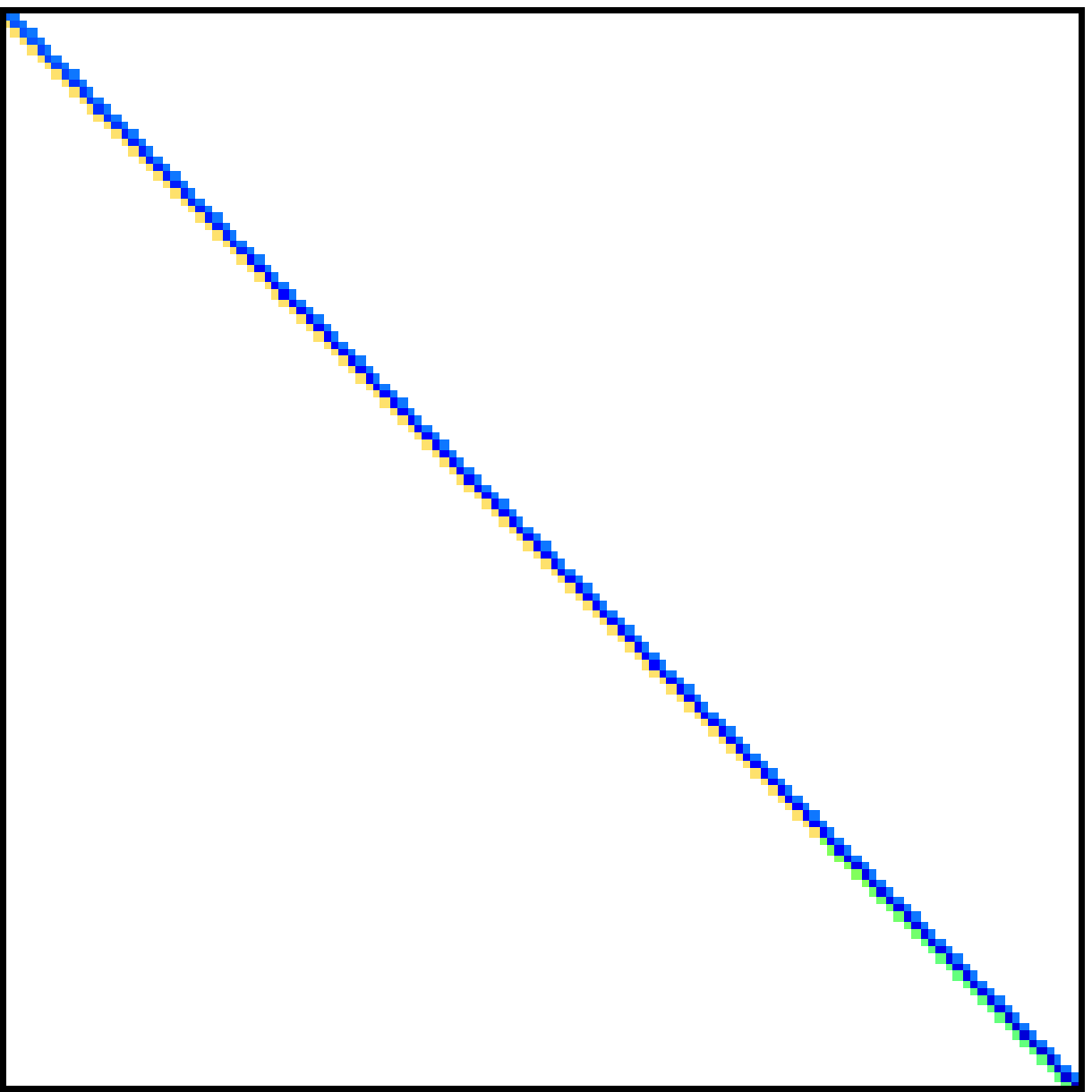} \end{minipage} \\ 
\textbf{Dimensions} & 
47 K $\times$ 47 K & 
49 K $\times$ 49 K  & 
141 K $\times$ 141 K & 
207 K $\times$ 207 K  & 
526 K $\times$ 526 K \\
\boldmath{$nnz$} & 
2.4 M & 
1.9 M & 
7.8 M & 
1.3 M & 
2.1 M \\
\begin{minipage}[c]{0.15\columnwidth} \centering \boldmath{$nnz/row$} \textbf{(min, avg, max)} \end{minipage}  & 
4, 50, 145 & 
39, 39, 39 & 
24, 55, 102 & 
1, 6, 44 & 
2, 3, 4 \\
\hline

\textbf{Name} & 
FEM/Accelerator  & 
Circuit & 
Webbase & 
LP & 
ASIC\_680k  \\
\textbf{Plot} &
\begin{minipage}[c]{0.12\columnwidth} \centering \includegraphics[width=0.6in]{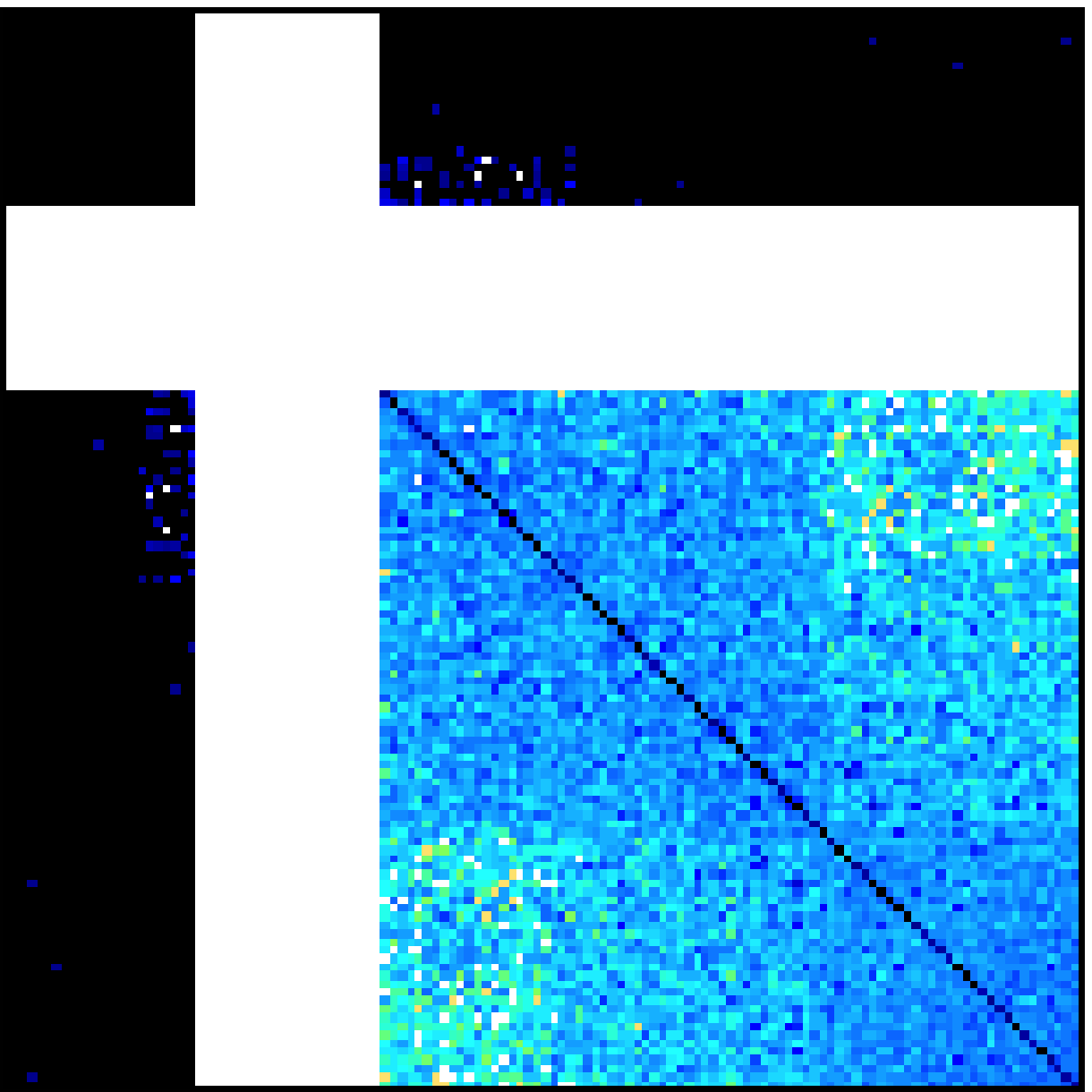} \end{minipage} &
\begin{minipage}[c]{0.12\columnwidth} \centering \includegraphics[width=0.6in]{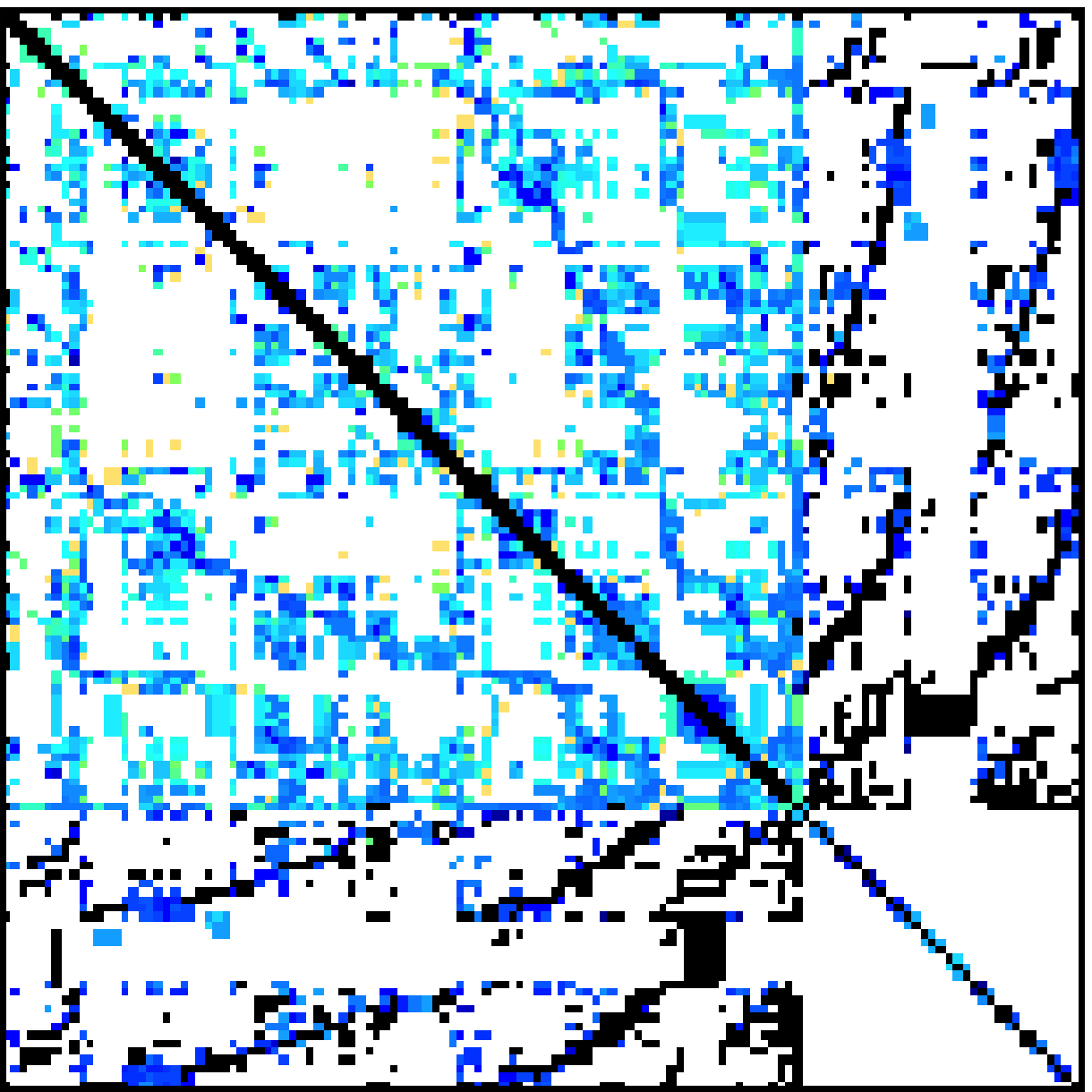} \end{minipage} &
\begin{minipage}[c]{0.12\columnwidth} \centering \includegraphics[width=0.6in]{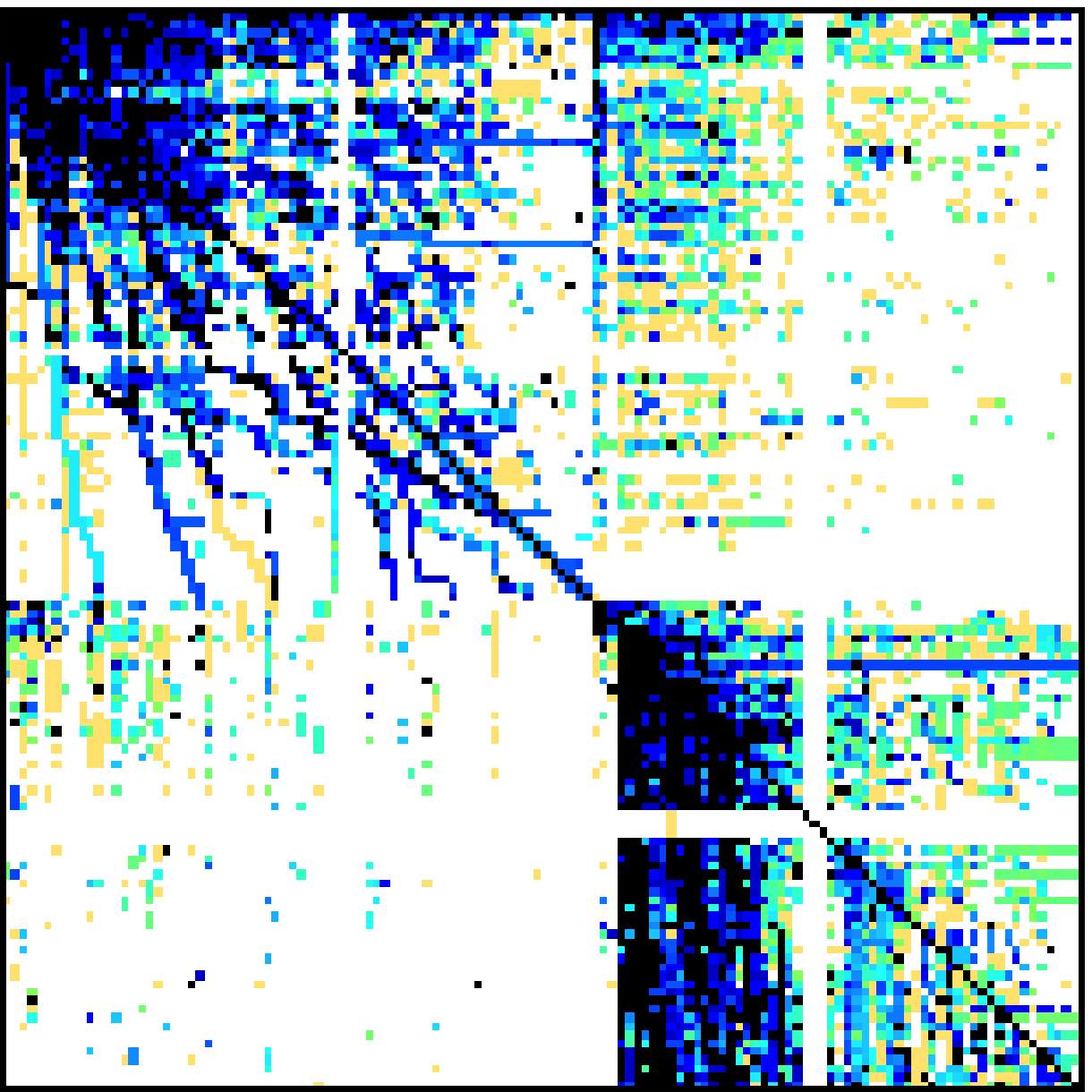} \end{minipage} &
\begin{minipage}[c]{0.12\columnwidth} \centering \includegraphics[width=0.6in]{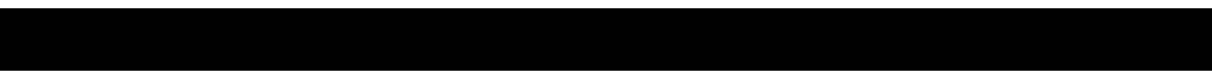} \end{minipage} &
\begin{minipage}[c]{0.12\columnwidth} \centering \includegraphics[width=0.6in]{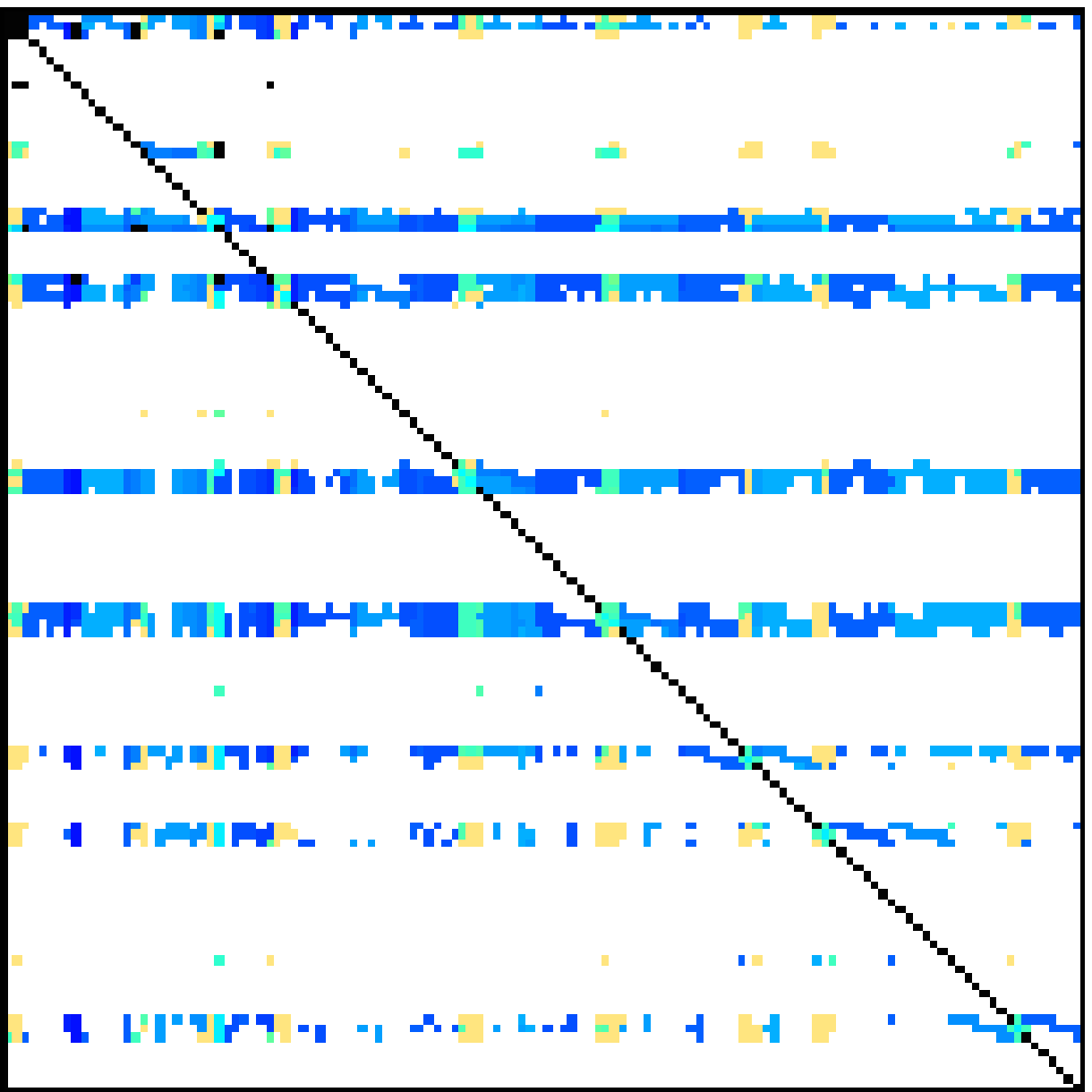} \end{minipage} \\ 
\textbf{Dimensions} & 
121 K $\times$ 121 K  & 
171 K $\times$ 171 K & 
1 M $\times$ 1 M & 
4 K $\times$ 1.1 M & 
683 K $\times$ 683 K  \\
\boldmath{$nnz$} & 
2.6 M & 
959 K & 
3.1 M & 
11.3 M & 
3.9 M \\
\begin{minipage}[c]{0.15\columnwidth} \centering \boldmath{$nnz/row$} \textbf{(min, avg, max)} \end{minipage}  & 
0, 21, 81 & 
1, 5, 353 & 
1, 3, 4.7 K & 
1, 2.6 K, 56.2 K & 
1, 6, 395 K \\
\hline

\textbf{Name} & 
boyd2 & 
dc2  & 
ins2  & 
rajat21 & 
transient \\
\textbf{Plot} &
\begin{minipage}[c]{0.12\columnwidth} \centering \includegraphics[width=0.6in]{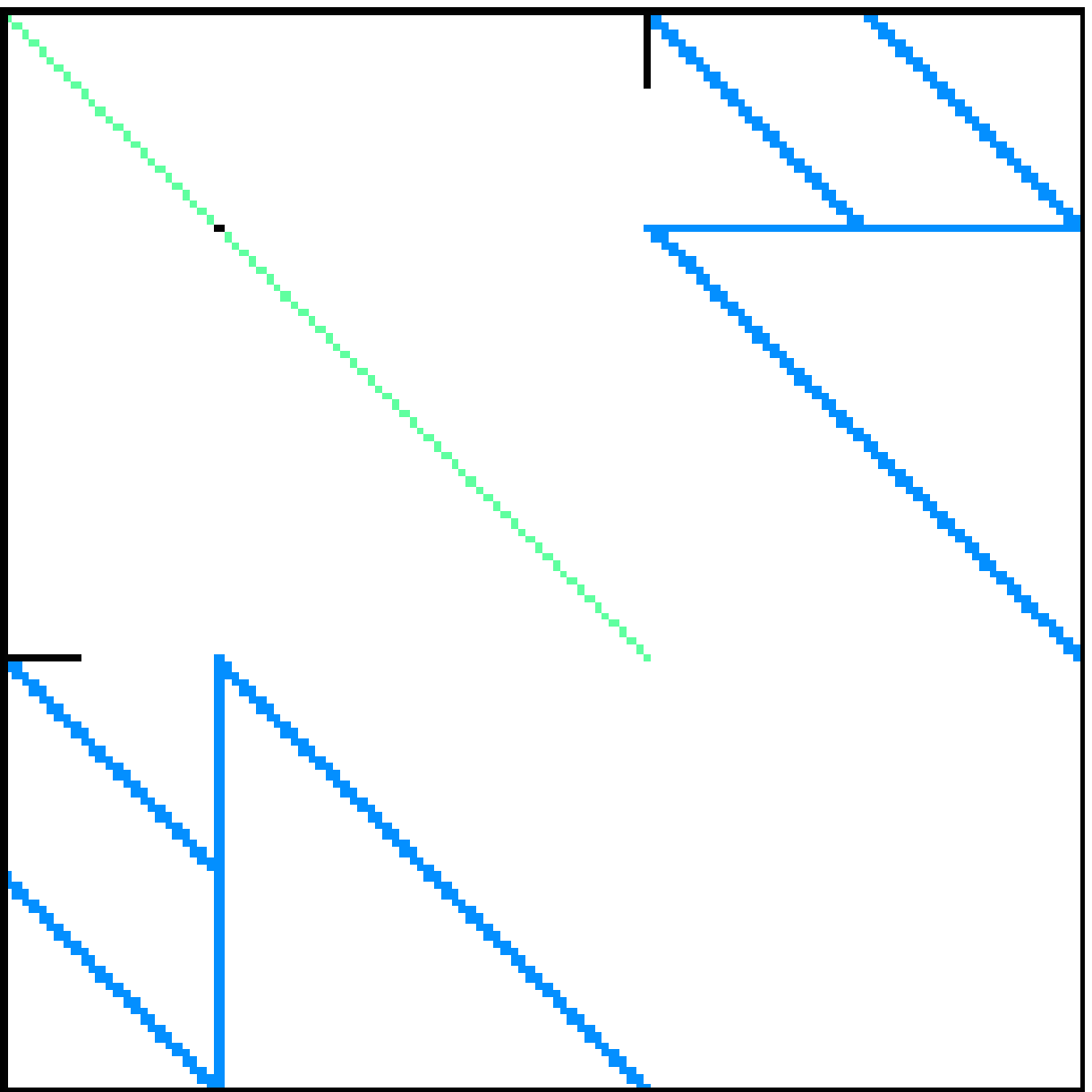} \end{minipage} &
\begin{minipage}[c]{0.12\columnwidth} \centering \includegraphics[width=0.6in]{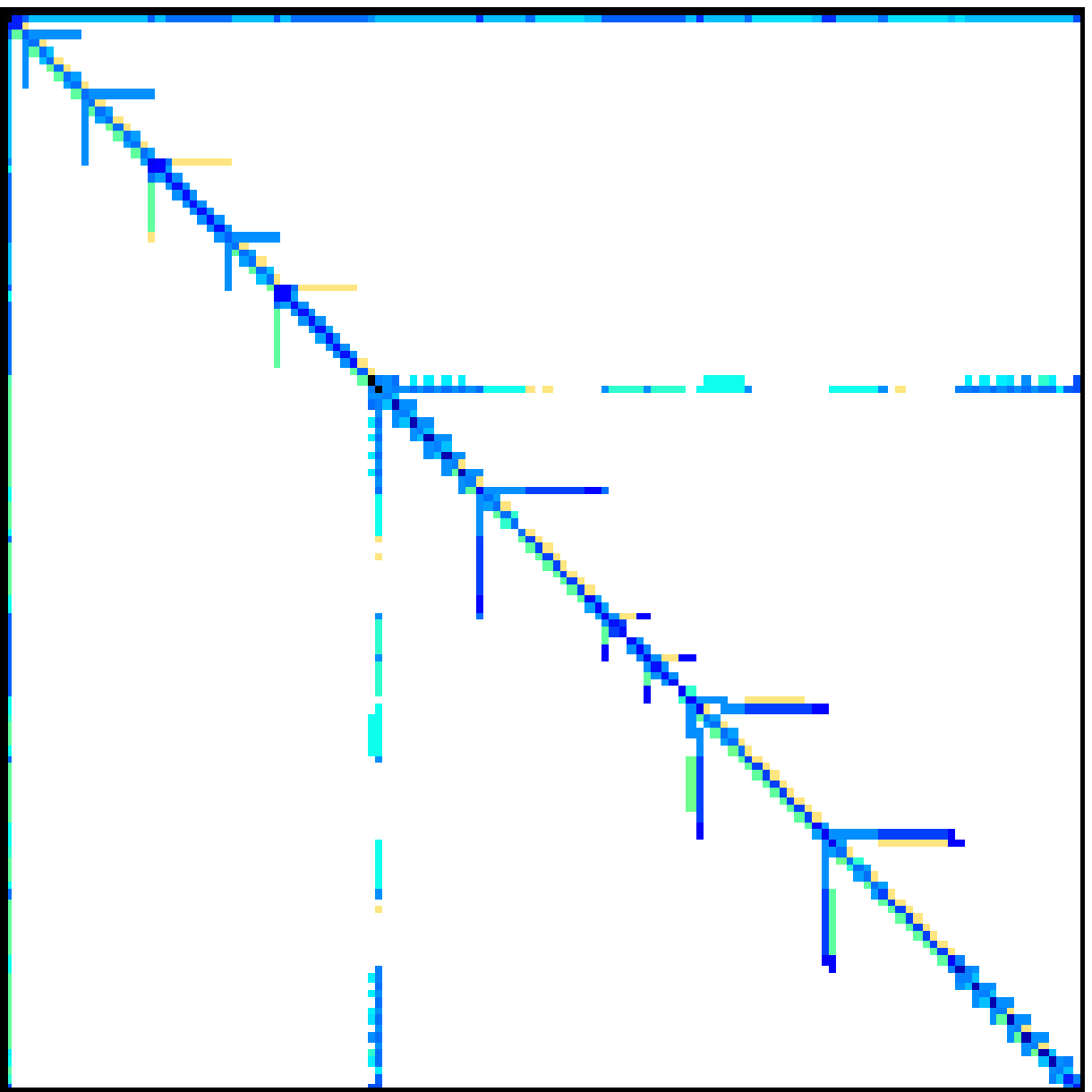} \end{minipage} &
\begin{minipage}[c]{0.12\columnwidth} \centering \includegraphics[width=0.6in]{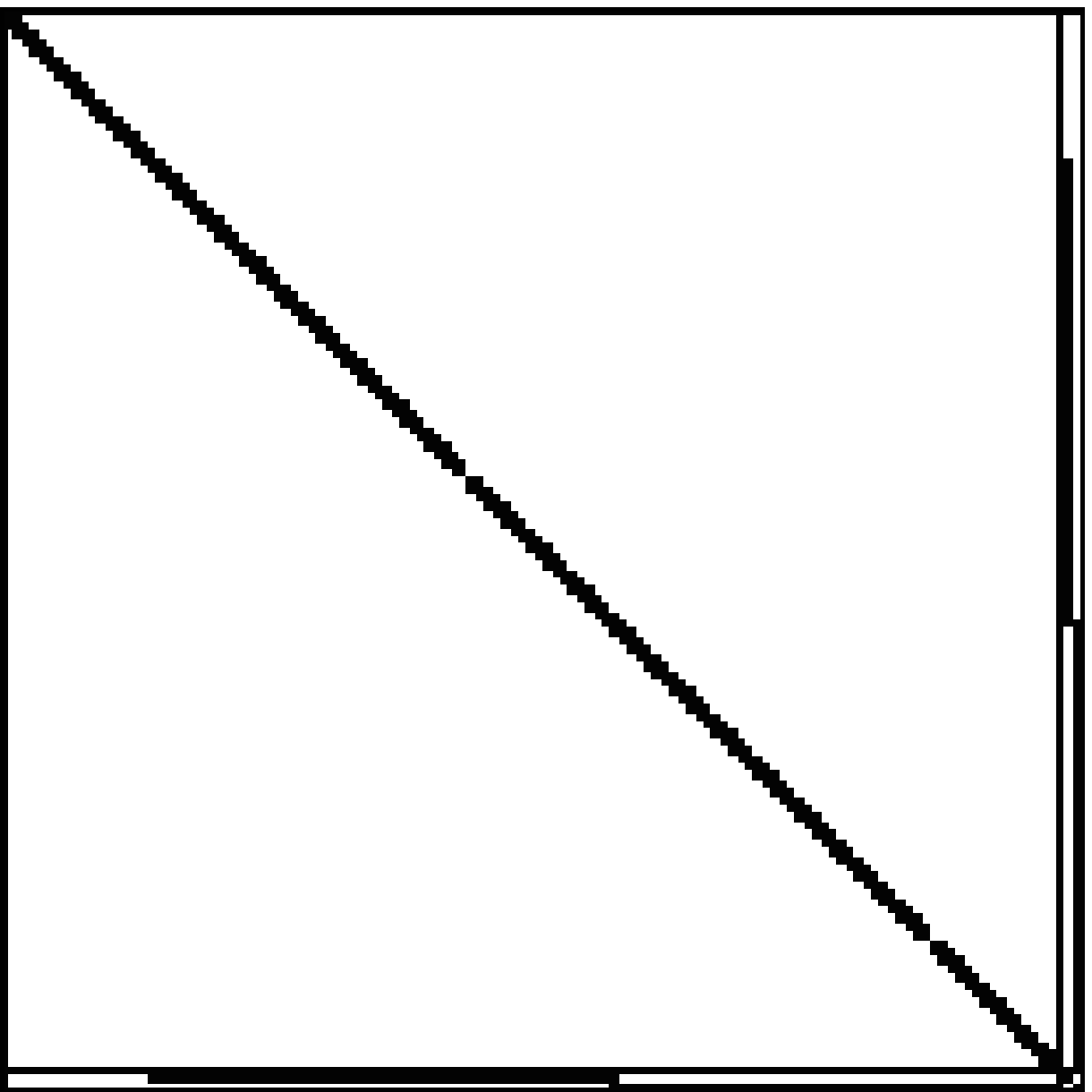} \end{minipage} &
\begin{minipage}[c]{0.12\columnwidth} \centering \includegraphics[width=0.6in]{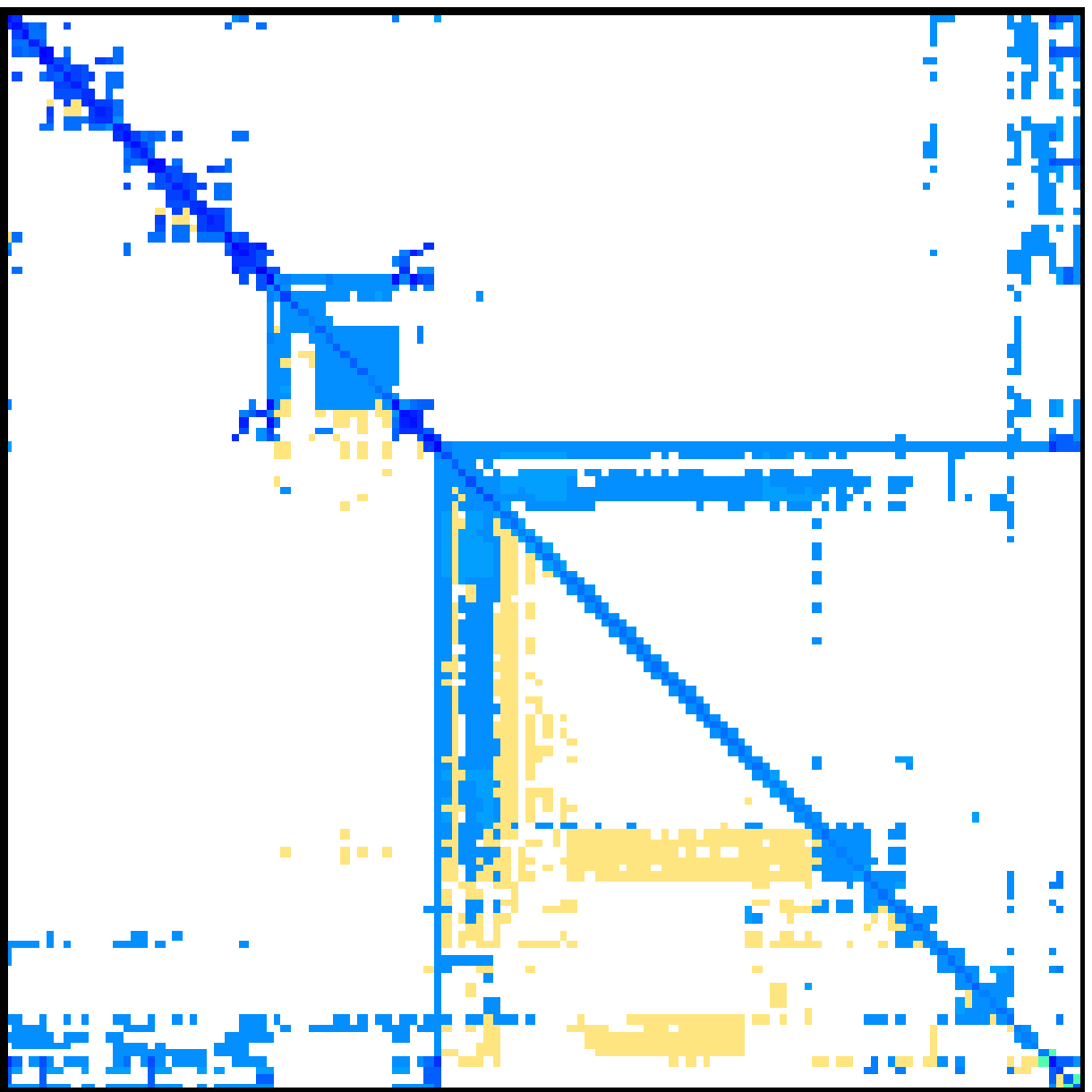} \end{minipage} &
\begin{minipage}[c]{0.12\columnwidth} \centering \includegraphics[width=0.6in]{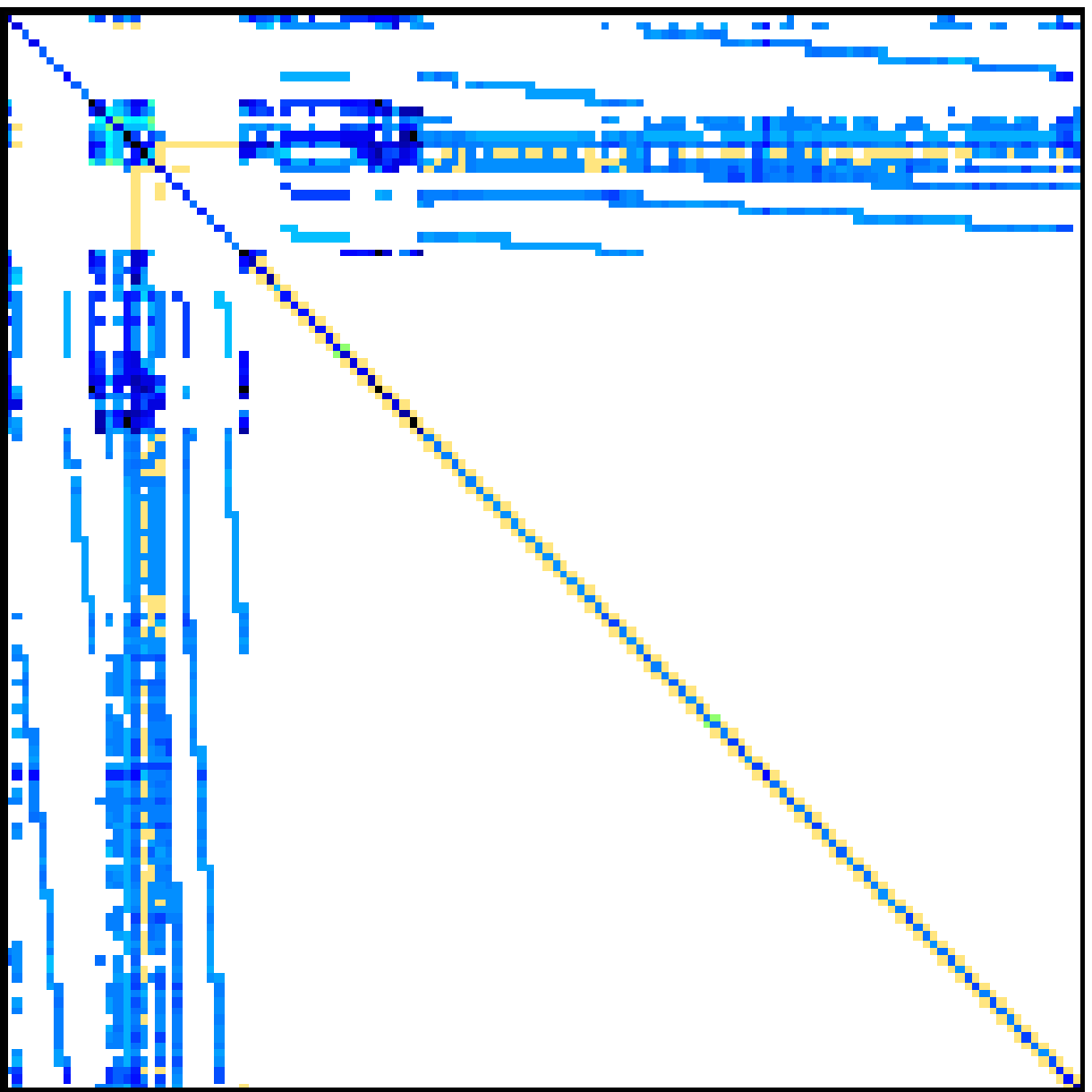} \end{minipage} \\ 
\textbf{Dimensions} & 
466 K $\times$ 466 K & 
117 K $\times$ 117 K & 
309 K $\times$ 309 K & 
412 K $\times$ 412 K & 
179 K $\times$ 179 K  \\
\boldmath{$nnz$} & 
1.5 M & 
766 K & 
2.8 M & 
1.9 M & 
962 K \\
\begin{minipage}[c]{0.15\columnwidth} \centering \boldmath{$nnz/row$} \textbf{(min, avg, max)} \end{minipage}  & 
2, 3, 93 K & 
1, 7, 114 K & 
5, 9, 309 K & 
1, 5, 119 K & 
1, 5, 60 K \\
\hline

\end{tabular}
\end{table}

\subsection{Experimental Setup}

To analyze efficiency of the proposed SpMV algorithm, we also benchmark parallel CSR-based SpMV using some other libraries or methods on CPUs and GPUs. 

On CPUs, we execute three CSR-based SpMV approaches: (1) OpenMP-accelerated basic row block method, (2) pOSKI library~\cite{Byun:pOSKI} using OSKI~\cite{Vuduc:OSKI} as a building block, and (3) Intel MKL v11.2 Update 2 in Intel Parallel Studio XE 2015 Update 2. The three approaches are running on all CPU cores of the used heterogeneous processors.  For the Intel CPU, we report results from MKL, since it always delivers the best performance and the pOSKI is not supported by the used Microsoft Windows operating system. For the AMD CPU, we report the best results of the three libraries, since none of the three libraries outperforms all the others. For the ARM CPU included in the nVidia Tegra K1 platform, we only report results from OpenMP, since the current pOSKI and Intel MKL implementations do not support the ARM architecture. Moreover, single-threaded na\"{\i}ve implementation on CPU is included in our benchmark as well.

On GPUs, we benchmark variants of the CSR-scalar and the CSR-vector algorithms proposed in~\cite{Bell:Implementing}. The OpenCL version of the CSR-scalar method is extracted from PARALLUTION v1.0.0~\cite{Lukarski:PARALUTION} and evaluated on the AMD platform. The OpenCL implementation of the CSR-vector method is extracted from semantically equivalent CUDA code in the CUSP library v0.4.0 and executed on both the Intel and the AMD platforms. On the nVidia platform, we run the CSR-based SpMV from vendor-supplied cuSPARSE v6.0 and CUSP v0.4.0 libraries.

For all tests, we run SpMV 200 times and record averages. The implicit data transfer (i.e., matrices and vectors data copy from their sources to OpenCL Shared Virtual Memory or CUDA Unified Memory) is not included in our evaluation, since SpMV operation is normally one building block of more complex applications. All participating methods conduct general SpMV, meaning that symmetry is not considered although some input matrices are symmetric. The throughput (flops per second) is calculated by 
\begin{equation} \label{eq1}
\begin{split}
\frac{2\times nnz}{runtime}. \nonumber
\end{split}
\end{equation}
The bandwidth (bytes per second) is calculated by 
\begin{equation} \label{eq1}
\begin{split}
\frac{(m + 1 + nnz) \times sizeof(idx\_type) + (nnz+nnz+m) * sizeof(val\_type)}{runtime}. \nonumber
\end{split}
\end{equation}

\subsection{Performance Analysis}

\begin{figure}[!t]
\captionsetup[subfigure]{labelformat=empty}
\centering
\subfloat[(a) Dense]{\epsfig{file=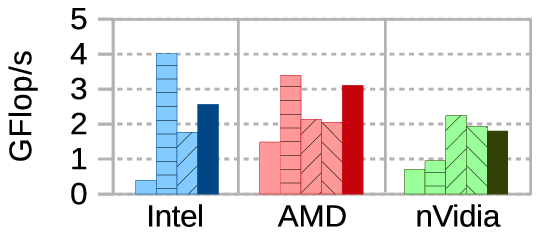, width=1.35in}}
\subfloat[(b) Protein]{\epsfig{file=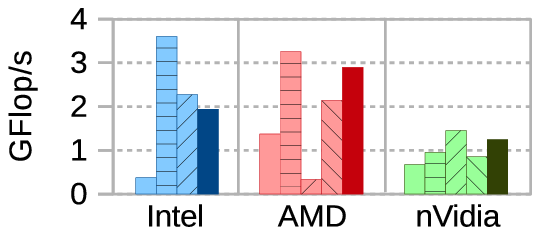, width=1.35in}}
\subfloat[(c) FEM/Spheres]{\epsfig{file=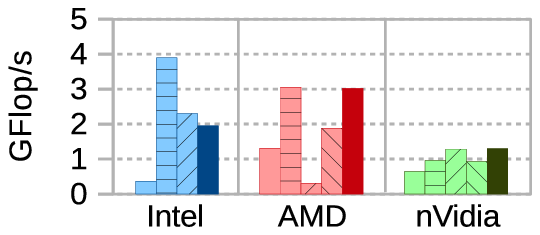, width=1.35in}}
\subfloat[(d) FEM/Cantilever]{\epsfig{file=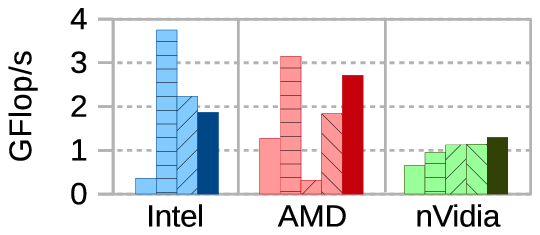, width=1.35in}}
\vskip -6pt
\qquad
\subfloat[(e) Wind Tunnel]{\epsfig{file=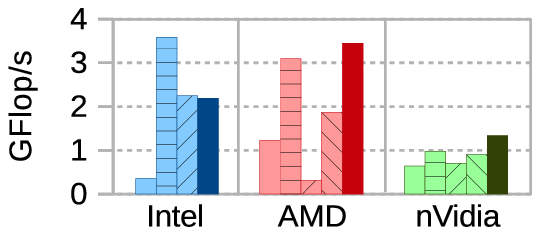, width=1.35in}}
\subfloat[(f) FEM/Harbor]{\epsfig{file=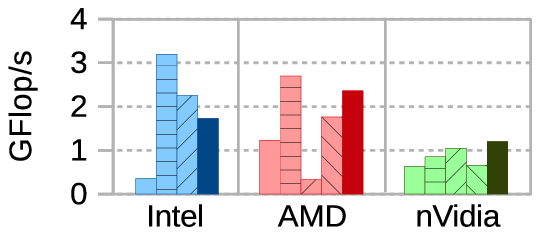, width=1.35in}}
\subfloat[(g) QCD]{\epsfig{file=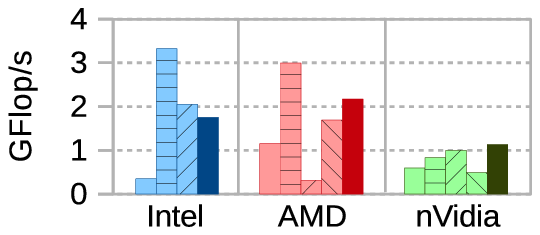, width=1.35in}}
\subfloat[(h) FEM/Ship]{\epsfig{file=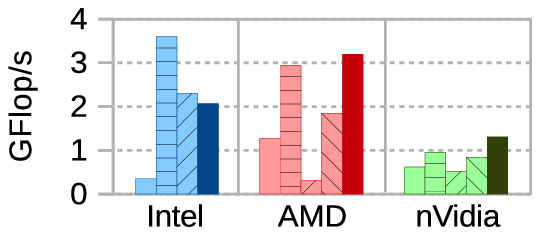, width=1.35in}}
\vskip -6pt
\qquad
\subfloat[(i) Economics]{\epsfig{file=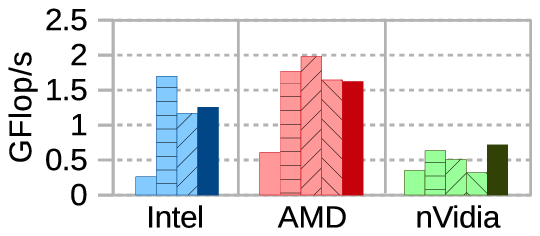, width=1.35in}}
\subfloat[(j) Epidemiology]{\epsfig{file=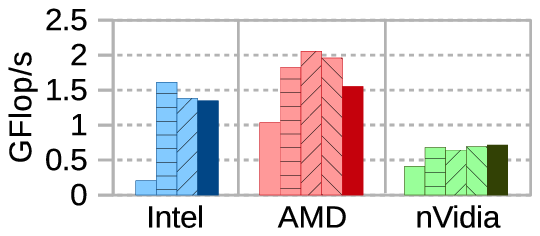, width=1.35in}}
\subfloat[(k) FEM/Accelerator]{\epsfig{file=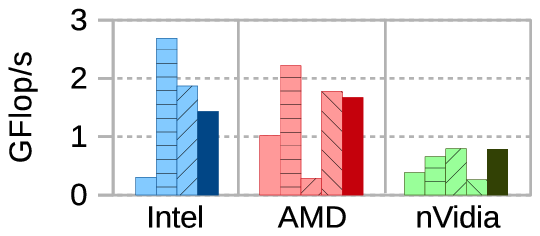, width=1.35in}}
\subfloat[(l) Circuit]{\epsfig{file=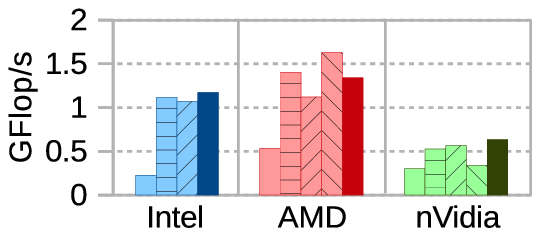, width=1.35in}}
\vskip -6pt
\qquad
\subfloat[(m) Webbase]{\epsfig{file=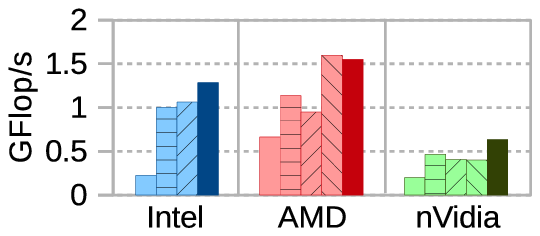, width=1.35in}}
\subfloat[(n) LP]{\epsfig{file=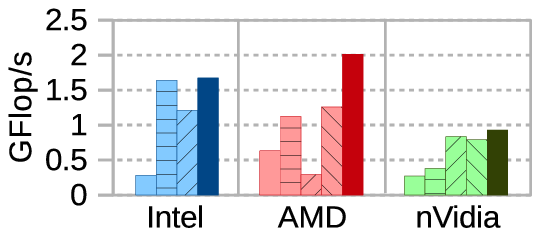, width=1.35in}}
\subfloat[(o) ASIC\_680k]{\epsfig{file=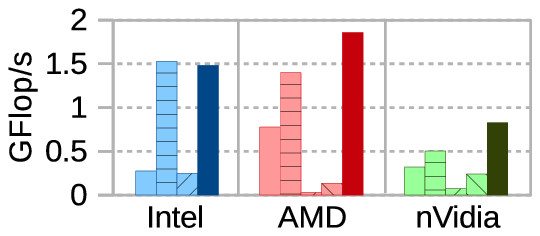, width=1.35in}}
\subfloat[(p) boyd2]{\epsfig{file=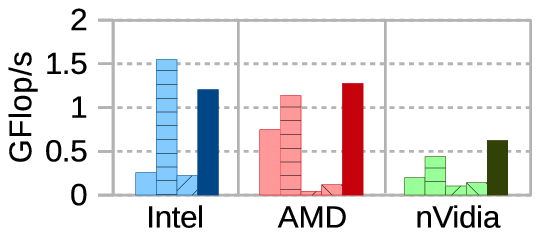, width=1.35in}}
\vskip -6pt
\qquad
\subfloat[(q) dc2]{\epsfig{file=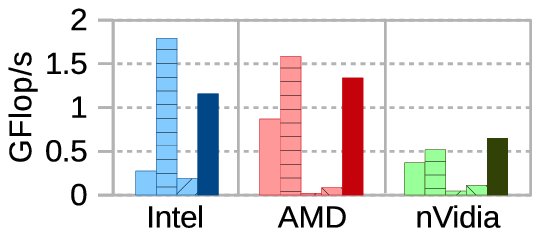, width=1.35in}}
\subfloat[(r) ins2]{\epsfig{file=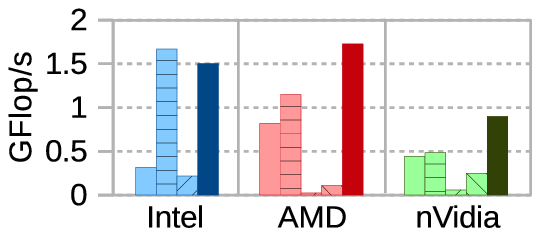, width=1.35in}}
\subfloat[(s) rajat21]{\epsfig{file=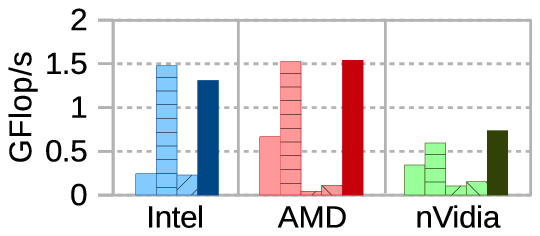, width=1.35in}}
\subfloat[(t) transient]{\epsfig{file=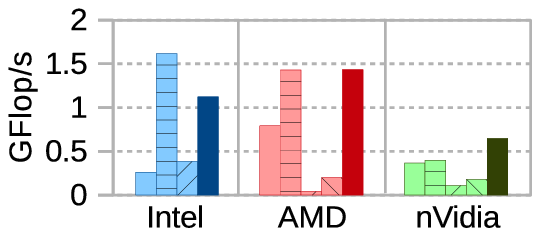, width=1.35in}}
\vspace{3mm}
\qquad
\subfloat[]{\epsfig{file=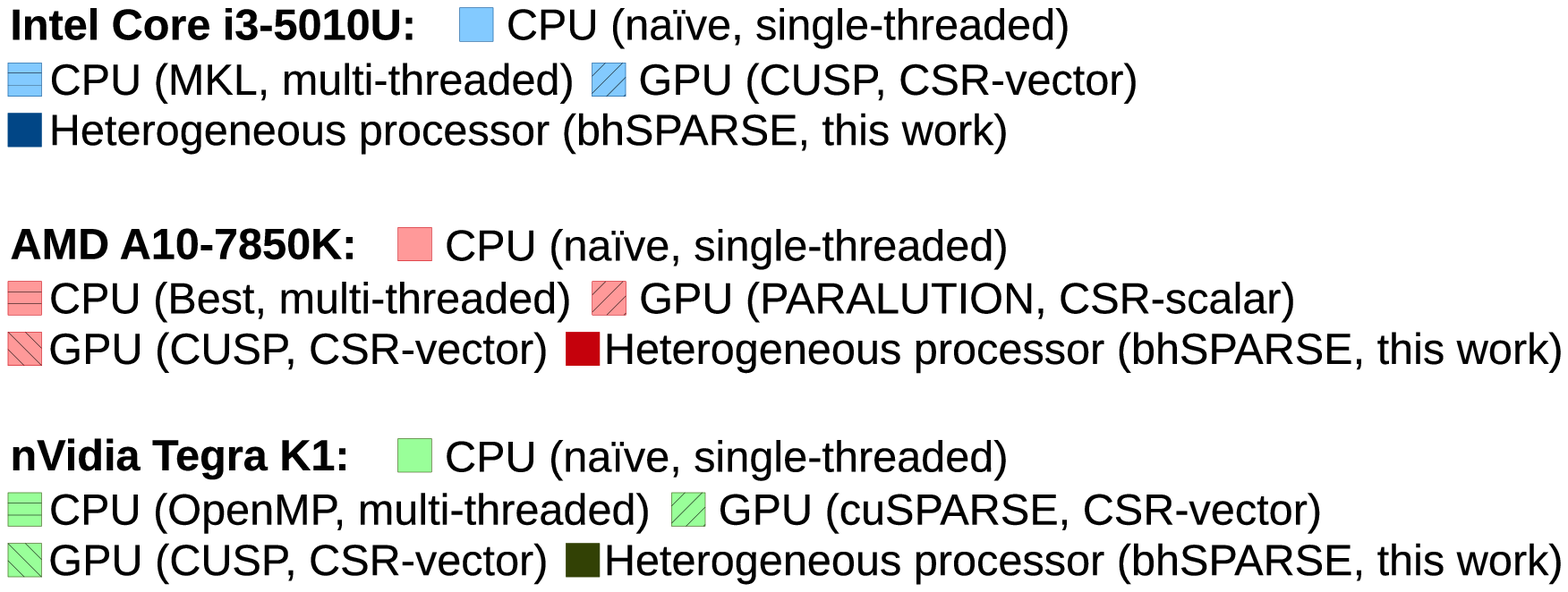, trim=0in 4.1in 0in 0in, width=2.9in}}
\subfloat[Harmonic mean]{\epsfig{file=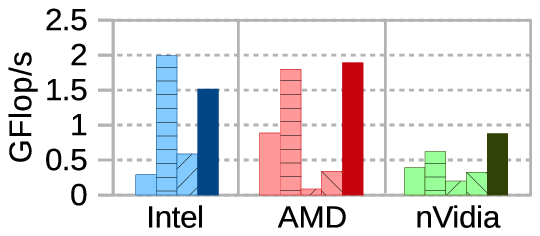, width=2.35in}}
\caption{Throughput (GFlop/s) of the single precision CSR-based SpMV algorithms running on the three platforms. ``CPU (Best, multi-threaded)'' shows the best results of OpenMP parallelization, pOSKI and Intel MKL on the AMD device. ``CSR-scalar'' and ``CSR-vector'' are variants of the row block algorithm on GPUs. ``bhSPARSE'' shows our CSR-based SpMV approach described in this paper.}
\label{spmv.parco.fig.spthroughput}
\end{figure}

\begin{figure}[!t]
\captionsetup[subfigure]{labelformat=empty}
\centering
\subfloat[(a) Dense]{\epsfig{file=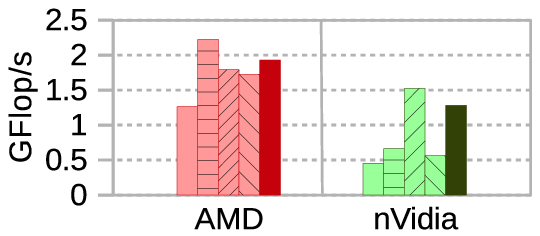, width=1.35in}}
\subfloat[(b) Protein]{\epsfig{file=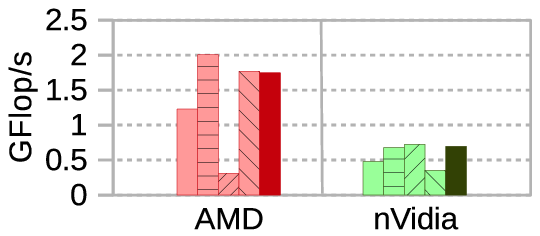, width=1.35in}}
\subfloat[(c) FEM/Spheres]{\epsfig{file=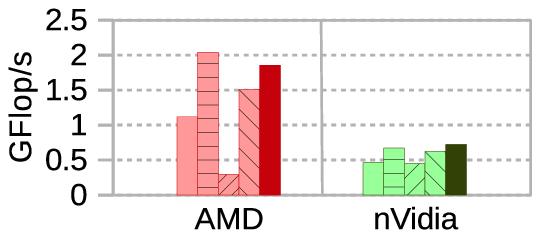, width=1.35in}}
\subfloat[(d) FEM/Cantilever]{\epsfig{file=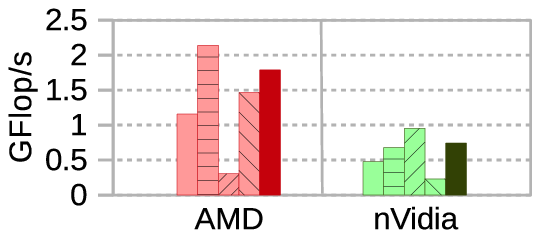, width=1.35in}}
\vskip -6pt
\qquad
\subfloat[(e) Wind Tunnel]{\epsfig{file=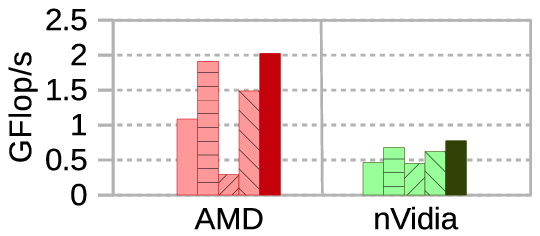, width=1.35in}}
\subfloat[(f) FEM/Harbor]{\epsfig{file=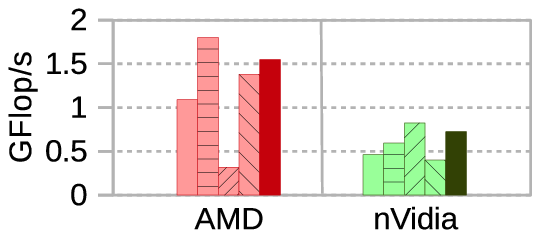, width=1.35in}}
\subfloat[(g) QCD]{\epsfig{file=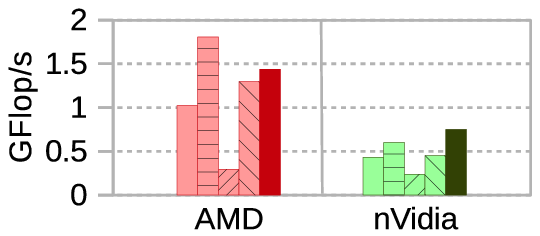, width=1.35in}}
\subfloat[(h) FEM/Ship]{\epsfig{file=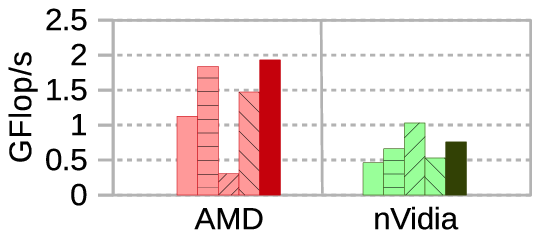, width=1.35in}}
\vskip -6pt
\qquad
\subfloat[(i) Economics]{\epsfig{file=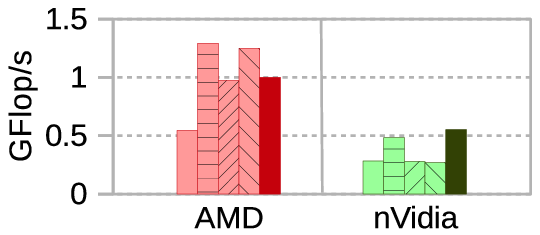, width=1.35in}}
\subfloat[(j) Epidemiology]{\epsfig{file=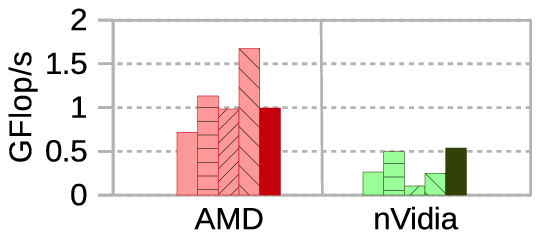, width=1.35in}}
\subfloat[(k) FEM/Accelerator]{\epsfig{file=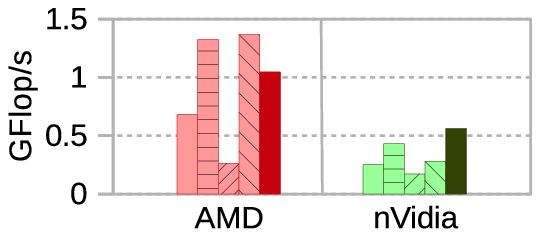, width=1.35in}}
\subfloat[(l) Circuit]{\epsfig{file=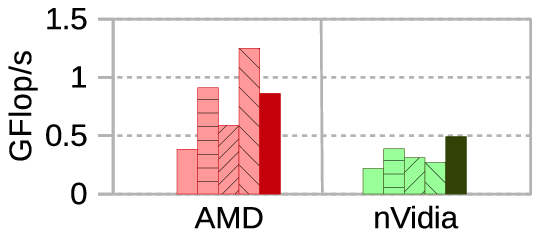, width=1.35in}}
\vskip -6pt
\qquad
\subfloat[(m) Webbase]{\epsfig{file=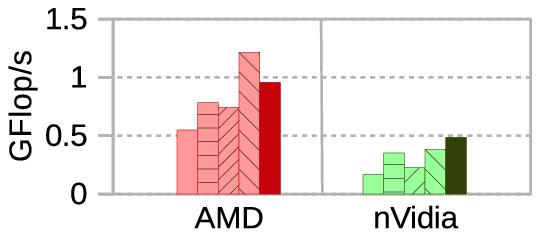, width=1.35in}}
\subfloat[(n) LP]{\epsfig{file=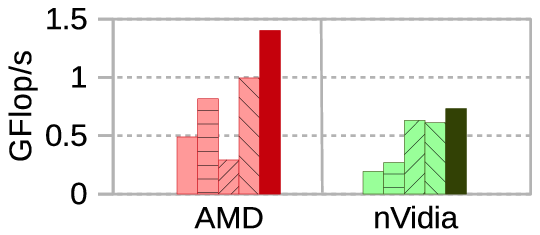, width=1.35in}}
\subfloat[(o) ASIC\_680k]{\epsfig{file=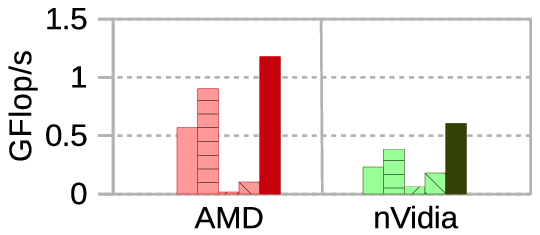, width=1.35in}}
\subfloat[(p) boyd2]{\epsfig{file=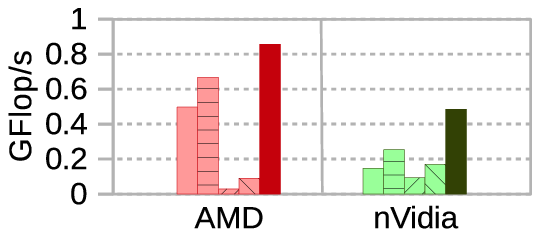, width=1.35in}}
\vskip -6pt
\qquad
\subfloat[(q) dc2]{\epsfig{file=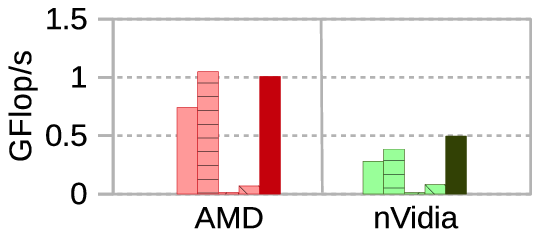, width=1.35in}}
\subfloat[(r) ins2]{\epsfig{file=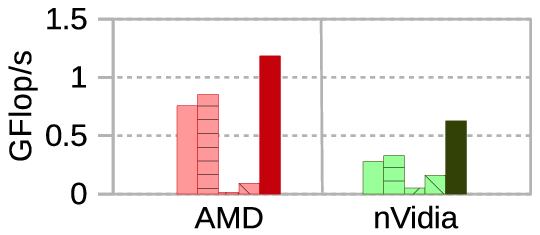, width=1.35in}}
\subfloat[(s) rajat21]{\epsfig{file=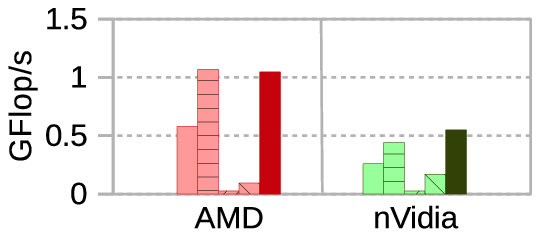, width=1.35in}}
\subfloat[(t) transient]{\epsfig{file=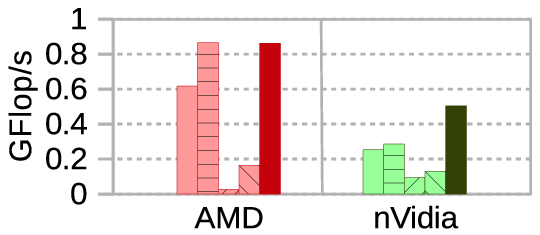, width=1.35in}}
\vspace{3mm}
\qquad
\subfloat[]{\epsfig{file=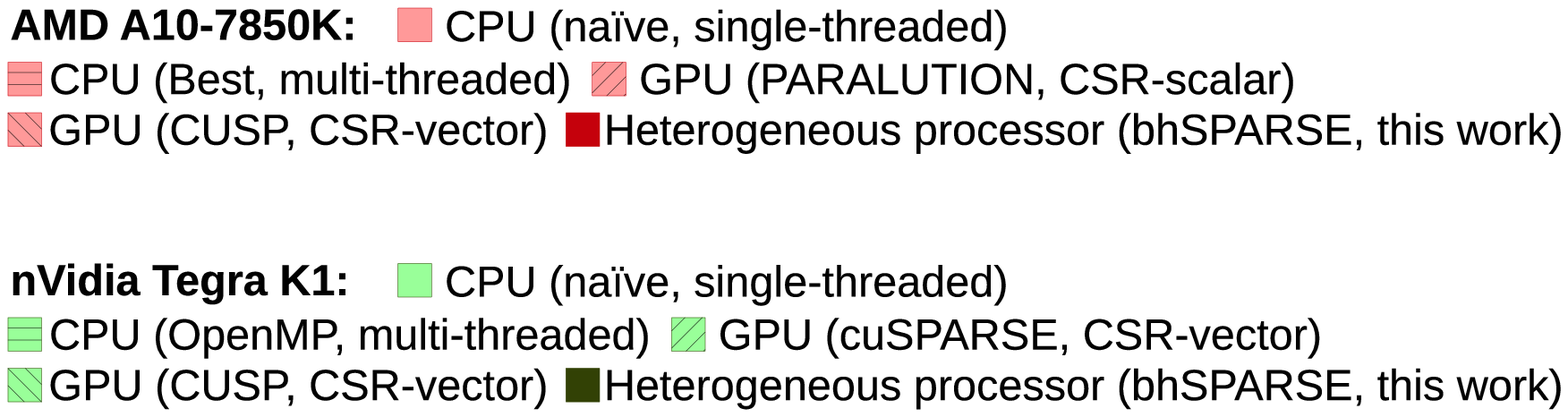, trim=0in 4.1in 0in 0in, width=2.9in}}
\subfloat[Harmonic mean]{\epsfig{file=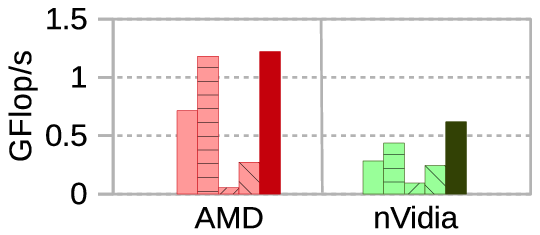, width=2.35in}}
\caption{Throughput (GFlop/s) of the double precision CSR-based SpMV algorithms running on the AMD and the nVidia platforms.}
\label{spmv.parco.fig.dpthroughput}
\end{figure}

Figures~\ref{spmv.parco.fig.spthroughput} and~\ref{spmv.parco.fig.dpthroughput} show throughput of single precision and double precision SpMV of the tested CSR-based approaches, respectively. 

In Figure~\ref{spmv.parco.fig.spthroughput}, we can see that on the Intel heterogeneous processor, our approach obtains up to 6.90x and on average 2.57x speedup over the CSR-vector method running on the used GPU. Although the speedup mainly comes from irregular matrices, our method generally does not obviously lose performance on regular matrices. Further, compared to CPU cores running MKL, both GPU SpMV algorithms are slower. For our algorithm, the main reason is that the integrated GPU implements scratchpad memory in its L3 cache, which has one order of magnitude higher latency compared to fast scratchpad in nVidia or AMD GPUs. Our algorithm in fact heavily uses scratchpad memory for storing and reusing segment descriptor, element-wise products and other shared data by threads. Thus even though the GPU part of the Intel heterogeneous processor has higher single precision theoretical peak performance than its CPU part, the delivered SpMV throughput is lower than expected. For the CSR-vector method, the low performance has another reason: small thread-bunch of size 8 dramatically increases loop overhead~\cite{Baskaran:A}, which is one of the well known bottlenecks~\cite{Fang:A} of GPU programming.
 
In Figures~\ref{spmv.parco.fig.spthroughput} and~\ref{spmv.parco.fig.dpthroughput}, we can see that on the AMD heterogeneous processor, our method delivers up to 71.90x (94.05x) and on average 22.17x (22.88x) speedup over the single (double) precision CSR-scalar method running on the used GPU. Compared to the GPU CSR-vector method, our algorithm achieves up to 16.07x (14.43x) and on average 5.61x (4.47x) speedup. The CSR-scalar and the CSR-vector methods give very low throughput while running the last 6 irregular matrices, because of the problem of load imbalance. Further, we find that the Intel heterogeneous processor's GPU is actually faster than the AMD GPU while running the last 6 matrices. The reason is that the shorter thread-bunch (8 in Intel GPU vs. 64 in AMD GPU) brings a positive influence for saving SIMD idle cost by executing a much shorter vector width for dramatically imbalanced row distribution. On the other hand, for several very regular matrices with short rows, e.g., \textit{Epidemiology}, the CSR-scalar method offers the best performance because of almost perfect load balance and execution of short rows without loop cost. For most regular matrices, our method delivers comparable performance over the best CPU algorithm. 


In Figures~\ref{spmv.parco.fig.spthroughput} and~\ref{spmv.parco.fig.dpthroughput}, we can see that on the nVidia platform, our method delivers up to 5.91x (6.20x) and on average 2.69x (2.53x) speedup over the single (double) precision SpMV in the CUSP library running on the used GPU. Compared to cuSPARSE, our method has higher speedups. Since the both libraries use CSR-vector algorithm, those speedups are within expectations. Consider the Tegra K1 platform only contains one single GPU core, the problem of load imbalance on this device is not as heavy as on the above AMD platform. As a result, the speedups are not as high as those from the AMD processor. Here our method delivers on average 1.41x (1.42x) speedup over OpenMP-accelerated SpMV on the quad-core ARM CPU, while using single (double) precision benchmark.



\begin{figure}[!ht]
\centering
\subfloat[Single precision SpMV on Intel Core i3-5010U]{\epsfig{file=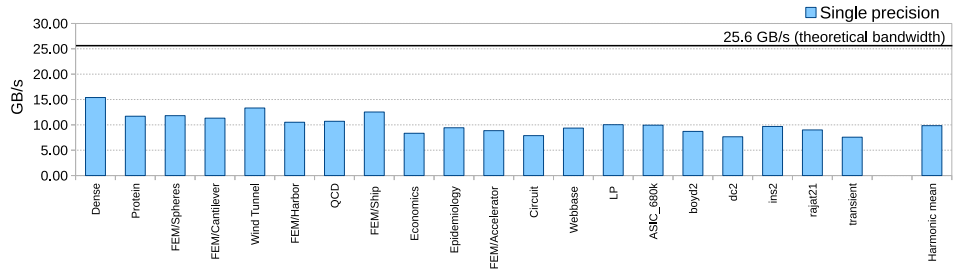, width=5in}}\qquad
\subfloat[Single precision and double precision SpMV on AMD A10-7850K]{\epsfig{file=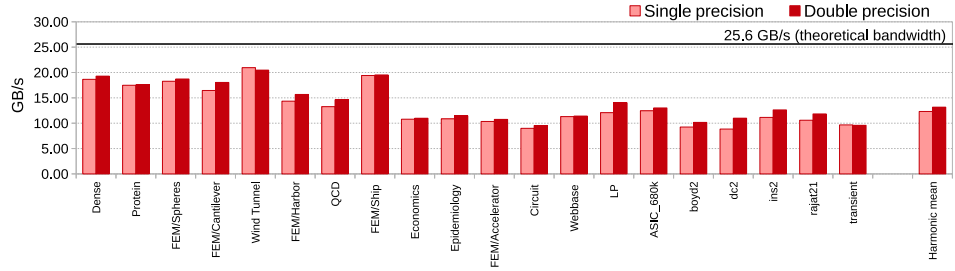, width=5in}}\qquad
\subfloat[Single precision and double precision SpMV on nVidia Tegra K1]{\epsfig{file=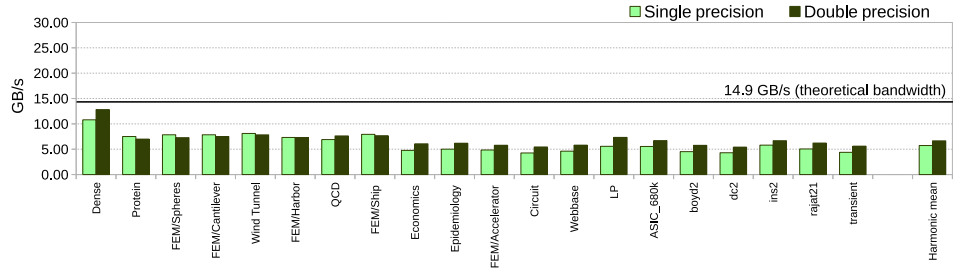, width=5in}}
\caption{Bandwidth utilization (GB/s) of our CSR-based SpMV algorithm running on the three platforms. Theoretical bandwidth from the hardware specifications are marked up using black lines.}
\label{spmv.parco.fig.bandwidth}
\end{figure}

Figure~\ref{spmv.parco.fig.bandwidth} shows bandwidth utilization of our algorithm proposed in this paper. We can see that the regular matrices can use bandwidth more efficiently compared to the irregular ones. Considering the throughput speedups listed above, our method can obtain higher bandwidth utilization than the other CSR-based SpMV algorithms running on GPUs. 

\subsection{Parameter Selection}

We further conduct experiments to exploit how selected parameters influence overall performance. 

\begin{figure}[!ht]
\centering
\subfloat[Intel, SP]{\epsfig{file=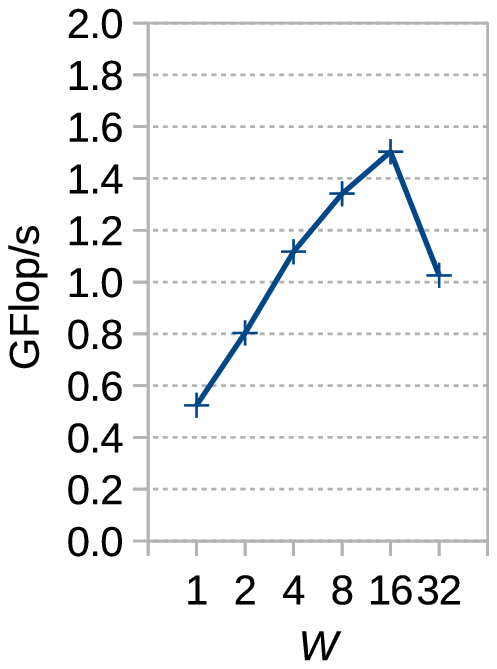, width=1.05in}}
\subfloat[AMD, SP]{\epsfig{file=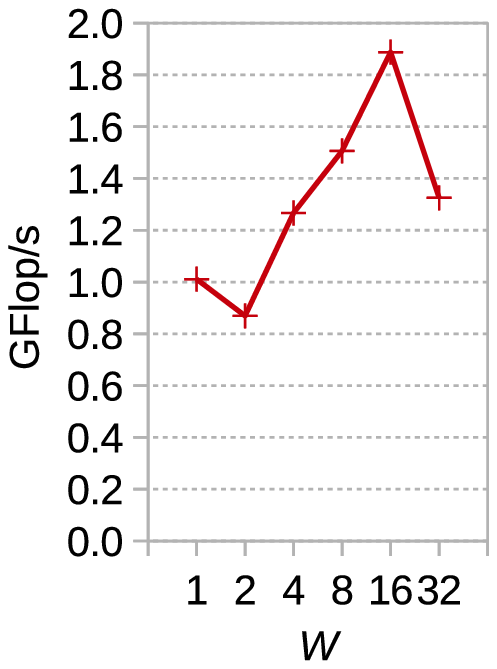, width=1.05in}}
\subfloat[nVidia, SP]{\epsfig{file=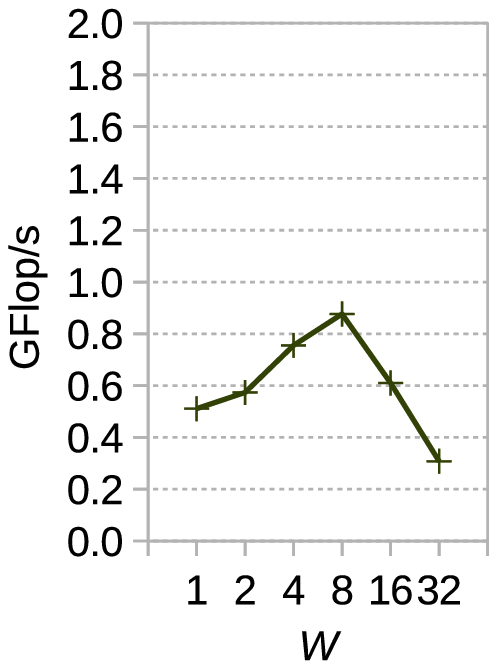, width=1.05in}}
\subfloat[AMD, DP]{\epsfig{file=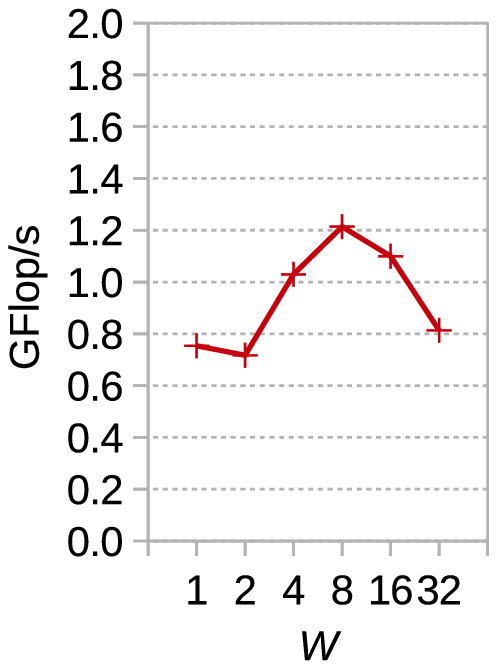, width=1.05in}}
\subfloat[nVidia, DP]{\epsfig{file=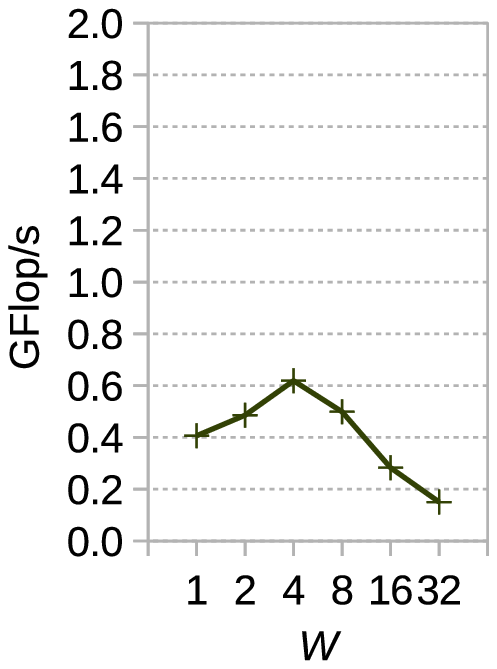, width=1.05in}}
\caption{Single precision (SP) and double precision (DP) SpMV performance of our algorithm on the three platforms while parameter $W$ changes and all the others fixed to the best observed values (see Table~\ref{spmv.parco.tab.parameters}).}
\label{spmv.parco.fig.parameterw}
\end{figure}

Figure~\ref{spmv.parco.fig.parameterw} shows dependency of the overall performance (harmonic means of the 20 benchmarks) on the parameters, while we fix all the parameters except for parameter $W$ (i.e., workload per thread). We can see that in general the overall performance goes up as parameter $W$ increases. This trend matches the algorithm complexity analysis described in Section 3.3. However, when $W$ is larger than a certain value, the overall performance degrades. The reason is that device occupancy may decrease while more on-chip scratchpad memory is allocated for $WT$ work space of each thread-bunch.

\begin{figure}[!ht]
\centering
\subfloat[Intel, SP]{\epsfig{file=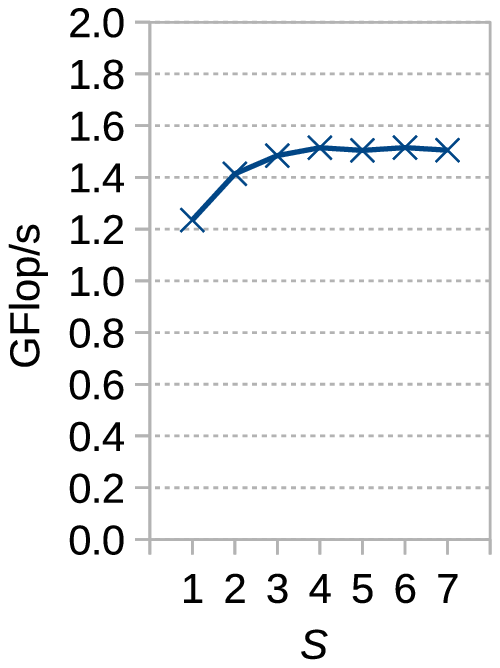, width=1.05in}}
\subfloat[AMD, SP]{\epsfig{file=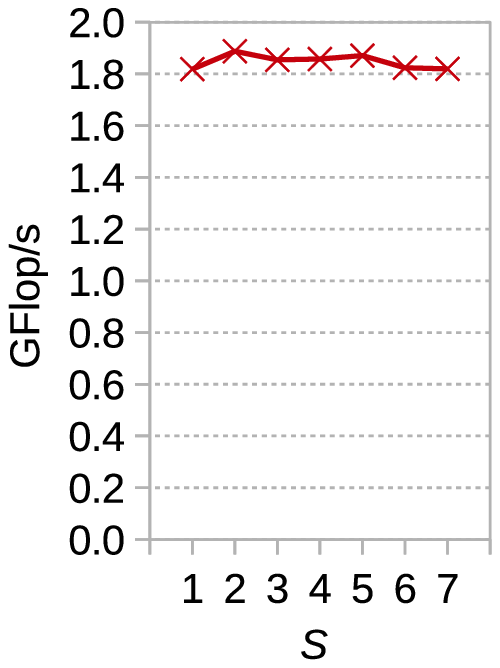, width=1.05in}}
\subfloat[nVidia, SP]{\epsfig{file=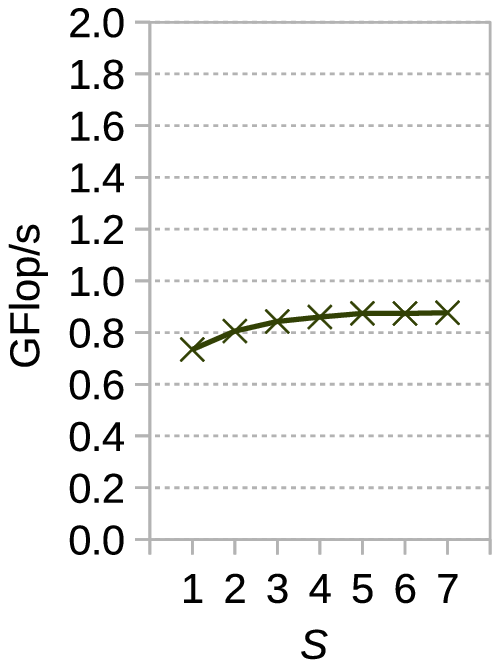, width=1.05in}}
\subfloat[AMD, DP]{\epsfig{file=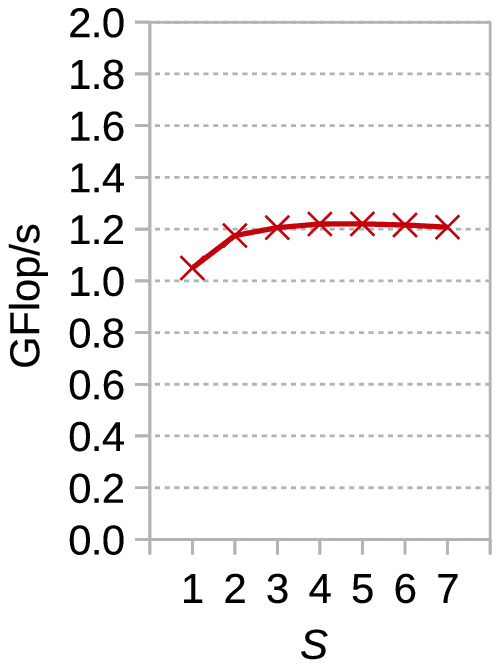, width=1.05in}}
\subfloat[nVidia, DP]{\epsfig{file=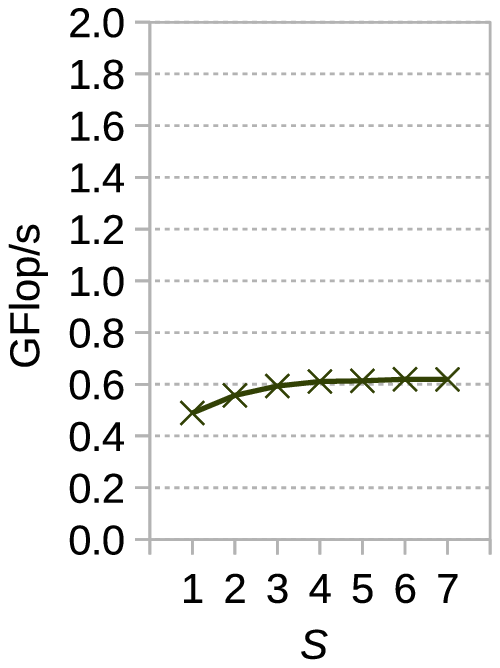, width=1.05in}}
\caption{Single precision (SP) and double precision (DP) SpMV performance of our algorithm on the three platforms while parameter $S$ changes and all the others fixed to the best observed values(see Table~\ref{spmv.parco.tab.parameters}).}
\label{spmv.parco.fig.parameters}
\end{figure}

Figure~\ref{spmv.parco.fig.parameters} shows the trend of the overall performance while we change parameter $S$ (i.e., the number of iterations of each thread-bunch) and fix all the other parameters. We can see that if we assign more work to each thread-bunch, a better performance can be expected. The performance improvement mainly comes from higher on-chip resource reuse.

\section{Comparison to Related Methods}



In recent years, some new formats have been designed for SpMV operation on various processor architectures. Because of less off-chip memory access and better on-chip memory localization, block-based formats or libraries, such as OSKI~\cite{Vuduc:OSKI, Vuduc:Fast, Vuduc:Automatic}, pOSKI~\cite{Byun:pOSKI},  CSB~\cite{Buluc:Parallel, Buluc:Reduced}, BELLPACK~\cite{Choi:Model}, BCCOO/BCCOO+~\cite{Yan:yaSpMV}, BRC~\cite{Ashari:An} and RSB~\cite{Martone:Efficient}, attracted the most attention. However, block-based formats heavily rely on sparsity structure, meaning that the input matrix is required to have a block structure to meet potential block layout. Therefore, block-based formats are mainly suitable for some matrices generated from scientific computation problems, but may not fit irregular matrices generated from graph applications. Our method proposed in this paper is insensitive to the sparsity structure of input matrix, thus a generally better performance is achieved.

A lot of research has focused on improving row block method CSR-based SpMV. Williams et al.~\cite{Williams:Optimization} proposed multiple optimization techniques for SpMV on multi-core CPUs and Cell B.E. processor. Nishtala et al.~\cite{Nishtala:When} designed a high-level data partitioning method for SpMV to achieve better cache locality on multicore CPUs. Pichel et al.~\cite{Pichel:Optimization} evaluated how reordering techniques influence performance of SpMV on GPUs. Baskaran and Bordawekar~\cite{Baskaran:Optimizing} improved off-chip and on-chip memory access patterns of SpMV on GPUs. Reguly and Giles~\cite{Reguly:Efficient} improved thread cooperation for better GPU cache utilization. Ashari et al.~\cite{Ashari:Fast} utilized static reordering and the Dynamic Parallelism scheme offered by nVidia GPUs for fast SpMV operation. Greathouse et al.~\cite{Greathouse:Efficient} grouped contiguous rows for better runtime load balancing on GPUs. LightSpMV~\ref{Liu:LightSpMV} proposed to dynamically distribute matrix rows over warps in order for more balanced CSR-based SpMV without the requirement of generating auxiliary data structures, and implemented this approach using atomic operations and warp shuffle functions as the fundamental building blocks. However, again, the row block methods cannot achieve good performance for input matrix with dramatically imbalanced row distribution. In contrast, our method is independent with the sparsity structure of input matrix.

Using segmented sum as a building block is potentially a better generic method for the CSR-based SpMV. An early segmented sum method GPU SpMV was introduced by Sengupta et al.~\cite{Sengupta:Scan} and Garland~\cite{Garland:Sparse} and implemented in the cuDPP library~\cite{Harris:CUDPP}. But the cost of segmented sum and global memory access degrade overall SpMV performance. Zhang~\cite{Zhang:A} improved backward segmented scan for a better cache efficiency and implemented the CSR-based SpMV on multicore CPUs. Recently, nVidia's Modern GPU library~\cite{Baxter:Modern} implemented an improved reduction method, which has been used as a back-end of cuDPP. However, its performance still suffered by pre- and post-processing empty rows in global memory space. Our method, in contrast, uses scratchpad memory more efficiently and utilizes the two types of cores in a heterogeneous processor for better workload distribution.

Compared with our recent work CSR5~\cite{Liu:CSR5}, a format designed for cross-platform SpMV on CPUs, GPUs and Xeon Phi, the SpMV approach presented in this paper does not need to process any format conversion or generate any auxiliary data for the input CSR matrix. Consider the format conversion from the CSR to the CSR5 merely needs the cost of a few SpMV operations, the CSR5-based SpMV and the CSR-based SpMV can find their own application scenarios, such as solvers with different number of iterations.

\section{Conclusion and Future Work}
We proposed an efficient method for SpMV on heterogeneous processors using the CSR storage format. On three mainstream platforms from Intel, AMD and nVidia, our method greatly outperforms row block method CSR-based SpMV algorithms running on GPUs. The performance gain mainly comes from our newly developed speculative segmented sum strategy that efficiently utilizes different types of cores in a heterogeneous processor. 

In this work, we assign different task to the CPU part and the GPU part in one heterogeneous processor. However, the heaviest workload (stage 1 in our method) currently only runs on GPU cores, while the CPU cores may be idle. Obviously, it is possible to schedule tasks in the first stage on both CPU cores and GPU cores simultaneously for potentially higher throughput. However, a heterogeneity-aware scheduling strategy is beyond the scope of the SpMV algorithm focused in this paper. We refer the reader to~\cite{Lee:Transparent, Kaleem:Adaptive, Shen:An, Shen:Improving} for recent progress on utilizing both CPU cores and GPU cores in a heterogeneous environment.

\section*{Acknowledgments}

The authors would like to thank James Avery for his valuable feedback. The authors also thank the anonymous reviewers for their insightful suggestions and comments on this paper.







\newpage 

\appendix

\section{Pseudo Code}

\begin{algorithm}[!ht]
\caption{The SPMD pseudo code of a thread-bunch in speculative execution stage of the CSR-based SpMV}
\label{spmv.parco.alg.spmv}
\begin{algorithmic}[1]
    \Function{speculative\_execution\_gpu}{$ $}
      \State $tb \gets \Call{get-thread-globalid}{$ $} / T$;
    
      \State \emph{//positioning row indices of tiles in the thread-bunch}
      \For {$i=0$ to $S$} 
        \State \texttt{boundary[$i$]} $\gets tb \times S\times W\times T + i\times W\times T$
        \State \texttt{tile\_offset[$i$]}$\gets \Call{binary\_search}{\texttt{*row\_pointer}, \texttt{boundary[$i$]}}$
      \EndFor
      
      \State \emph{//iterative steps in a thread-bunch}
      \For {$i=0$ to $S-1$} 
        \State $start \gets \texttt{tile\_offset[$i$]}$
        \State $stop \gets \texttt{tile\_offset[$i+1$]}$
        
        \State \Call{memset}{$\texttt{*seg\_descriptor}, \texttt{FALSE}$}
        \State $dirty \gets \texttt{FALSE}$
        \State \emph{//calculating segment descriptor}
        \For {$j=start$ to $stop-1$} 
          \If {$\texttt{row\_pointer[$j$]} \neq \texttt{row\_pointer[$j+1$]}$}
            \State $\texttt{descriptor[row\_pointer[$j$] - boundary[$i$]]} \gets \texttt{TRUE}$
          \Else
            \State $dirty \gets \texttt{TRUE}$
          \EndIf
        \EndFor
        
        \State \emph{//collecting element-wise products}
        \For {$j=0$ to $W\times T-1$} 
          \State $x\_value \gets \texttt{x[column\_index[boundary[$i$]+$j$]]}$
          \State $\texttt{product[$j$]} \gets x\_value\times \texttt{value[boundary[$i$]+$j$]}$
        \EndFor
        
        \State \emph{//transmitting a value from the previous tile}
        \If {$\texttt{descriptor[$0$]} = \texttt{FALSE}$} 
          \State $\texttt{product[$0$]} \gets \texttt{product[0]} + transmitter$
          \State $\texttt{descriptor[$0$]} \gets \texttt{TRUE}$
        \EndIf
        
    \algstore{myalg}
    \end{algorithmic}
    \end{algorithm}

    \begin{algorithm}                     
    \begin{algorithmic} [1]                   
    \algrestore{myalg}
    
        \State \emph{//segmented sum}
        \State $\Call{segmented\_reduction}{\texttt{*product}, \texttt{*descriptor}, \texttt{*ts}, \texttt{*tc}}$ 
        \State \emph{//calculating index offset in $y$}
        \State $\texttt{*y\_index} \gets \Call{exclusive\_scan}{\texttt{*tc}}$ 

        \State \emph{//saving partial sums to $y$}
        \For {$j=0$ to $T-1$} 
          \For {$k=0$ to $\texttt{tc[$j$]}-1$}
            \State $index \gets start+\texttt{y\_index[$j$]}+k$
            \State \emph{//first segment of the thread-bunch}
            \If {$index = \texttt{tile\_offset[0]}$} 
              \State $\texttt{synchronizer[$tb$]}.idx \gets index$
              \State $\texttt{synchronizer[$tb$]}.val \gets \texttt{product[$j\times W+$ts[$j$]$+k$]}$

            \Else 
              \State \emph{//storing to $y$ directly}
              \State $\texttt{y[$index$]} \gets \texttt{product[$j\times W+$ts[$j$]$+k$]}$
            \EndIf
            
            \If {$index = stop$} 
              \State $transmitter \gets \texttt{product[$j\times W+$ts[$j$]$+k$]}$
            \EndIf
          \EndFor
        \EndFor

        \State \emph{//labeling dirty tile}
        \If {$dirty = \texttt{TRUE}$} 
          \State $pos \gets$ \Call{atomic\_increment}{$\texttt{*dirty\_counter}$}
          \State $\texttt{speculator[$pos$]} \gets \langle start, stop \rangle$
          \State $transmitter \gets 0$
        \EndIf
        
      \EndFor
    \EndFunction
    \State

    \Function{synchronization\_cpu}{$ $}
      \For {$i=0$ to $\lceil nnz/(S\times W\times T)\rceil-1$}
        \State $index \gets \texttt{synchronizer[$i$]}.idx$
        \State $value \gets \texttt{synchronizer[$i$]}.val$
        \State $\texttt{y[$index$]} \gets \texttt{y[$index$]} + value$
      \EndFor
    \EndFunction
  \end{algorithmic}
\end{algorithm}


\end{document}